\documentclass[final,3p,sort&compress,times]{elsarticle}
\usepackage{lineno,hyperref}
\usepackage{graphicx}
\usepackage{dcolumn}
\usepackage{bm}
\usepackage{upgreek}
\usepackage{amsmath}
\usepackage{multirow}
\usepackage{hyperref}

\usepackage{etoolbox}

\newcommand{\Nu}{N\!u}%
\newcommand{\VR}{V\!R}%
\journal{International Communications in Heat and Mass Transfer}

\begin{document}

\begin{frontmatter}

\title{Dynamics of vapor bubble train in flow boiling heat transfer in microchannels}

\author[SKLE]{Odumuyiwa A. Odumosu}
\author[SKLE,NIEPES]{Tianyou Wang}
\author[SKLE,NIEPES]{Zhizhao Che\corref{cor1}}
\cortext[cor1]{Corresponding author.}
\ead{chezhizhao@tju.edu.cn}
\address[SKLE]{State Key Laboratory of Engines, Tianjin University, Tianjin, 300350, China.}
\address[NIEPES]{National Industry-Education Platform of Energy Storage, Tianjin University, Tianjin, 300350, China}

\begin{abstract}
Microchannel flow boiling is a promising technique for micro-device thermal management, and understanding the bubble dynamics in microchannel flow boiling is important for the applications. Previous studies only focused on single, isolated bubbles, but the bubbles in microchannel flow boiling applications often exist as bubble trains, in which the bubbles interact with each other. Here, we investigate numerically vapor bubble trains in microchannel flow boiling by adopting the flow-focusing technique to form monodispersed bubbles in the upstream of the microchannel. With increasing the initial vapor-liquid volume ratio, the bubble frequency increases while the growth rate of the bubbles decreases because of the reduced bubble size. With increasing the heat flux on the wall or reducing the latent heat of the working fluid, the bubble train growth rate increases because of the increased vaporization rate. The vaporization of the fluid in the upstream causes the bubble expansion and accelerates the bubble movement in the downstream. The wall temperature and the Nusselt number fluctuate because of the periodic pass-through of bubbles. 
\end{abstract}

\begin{keyword}
\texttt {Bubble train \sep Boiling heat transfer \sep Bubble dynamics \sep Bubble interaction \sep Microchannel
}

\end{keyword}

\end{frontmatter}

\section{Introduction}\label{sec:1}
The rapid increase in heat dissipation of advanced microscale technologies requires effective heat removal techniques. Microchannel flow boiling has emerged as a promising technique for high-heat-flux removal in thermal management systems, with applications spanning from high-performance microelectronics \cite{karayiannis17, mudawar13} to concentrated photovoltaic cells \cite{valehesheyda13, wu24} and advanced nuclear reactors \cite{mudawar11}. A key phenomenon governing the heat transfer process of the flow boiling is the flow of bubble trains, where a series of vapor bubbles separated by liquid slugs flow through microchannels \cite{thome04}. Numerous studies have been carried out on the bubble behavior \cite{Huh2007FlowBoiling, yan23} in microchannel flow boiling to comprehend the complex intricacy of the boiling phenomenon, yet the process is not fully understood.

The key flow regimes in microchannel flow boiling include the single-phase liquid, bubbly, slug (confined bubble), churn, and annular flows \cite{harirchian11, thome15}. As the nucleated bubbles depart from the nucleation sites, they slide, disperse, grow, and coalesce, leading to large bubbles which rapidly fill the channel cross-section to form confined bubbles \cite{cheng17}. In slug flows, liquid slugs are separated from each other by vapor bubbles, and liquid films also exist between the vapor bubbles and the wall. With increased vaporization along the microchannel, the slug flow is changed into churn flow. Further downstream of the channel, annular flow occurs with a vapor core at the channel center with a liquid film on the wall \cite{kadam21}. The slug and annular regimes are considered the most dominant flow patterns \cite{Revellin2006DiabaticTwoPhaseFlow}. The rapid transition of the nucleate and the bubbly flow regimes to confined bubble and slug flow regimes at saturated conditions leads to continuous vapor bubble trains in the channel. 

The availability of advanced computing facilities with well robust numerical methods for two-phase flow has made simulations of boiling heat transfer possible \cite{Mukherjee2005BubbleGrowth}, serving as a vital tool to comprehend the flow boiling process in addition to experiments. Mukherjee et al.\ \cite{mukherjee11} performed numerical investigations on the bubble growth in microchannel flow boiling and found that the heat transfer is enhanced by the motion of the liquid-vapor interface during evaporation. Zhuan and Wang \cite{Zhuan2012FlowBoilingMicrochannel} numerically investigated the transitions of flow regimes and found that the flow transitions are the result of the bubbles' growth and subsequent coalescence. Ferrari et al.\ \cite{Ferrari2018SlugFlowBoiling} found that the channel geometry remarkably influences the heat transfer and because of the large density ratio of the liquid to vapor, the liquid evaporation makes the bubble travel faster. Luo et al.\ \cite{luo20} simulated the annular flow boiling, and revealed that the wall heat flux and the inlet vapor quality influence the thinness of the liquid film. Guo et al.\ \cite{guo16} simulated the annular flow regime and their results agree well with the correlations of film thickness. Priy et al.\ \cite{priy24} investigated numerically the nucleation and interaction of bubbles at nucleation sites during microchannel flow boiling and captured the vapor bubble generation, departure, sliding, growth, and coalescence within the channel. Zhang et al.\ \cite{zhang24} considered curved microchannels and found that the heat transfer is enhanced in curved channels compared to straight channels. Odumosu et al.\ \cite{odumosu23} considered wavy microchannels, and revealed that the perturbation of the flow field by wavy channel enhances the convection effect. Zhang et al.\ \cite{zhang23} considered vertical microchannels, and developed the phase diagram to predict the flow regime transitions. Rajkotwala et al.\ \cite{rajkotwala22} adopted the local front reconstruction method to simulate vapor bubbles in microchannel flow boiling, and achieved accurate tracking of the bubble interface. 

Despite many studies on microchannel flow boiling, most studies considered single isolated bubbles, and there are few studies on continuous bubble trains in microchannels. In bubble trains, the bubbles interact with each other via the flow and heat transfer, making the process more complex. Liu et al.\ \cite{Liu2017BubbleTrainMicrochannel} simulated the bubble dynamics and heat transfer of bubble trains in microchannel flow boiling, but the bubbles' flow is not continuous and only a few bubbles (i.e., 2--4) are considered. Their finding revealed that the local heat transfer efficiency is heavily affected by the liquid film thickness surrounding the bubbles. Magnini and Thome \cite{Magnini2016FlowBoiling} patched multiple vapor bubbles in the microchannel upstream with user-defined functions to mimic bubble stream during microchannel flow boiling and found that the flow dynamics of multiple bubbles are different compared to a single bubble scenario. The study of continuous flow of bubble trains is important because the heat transfer of microchannel heat sinks is heavily influenced by the dynamics of the bubble trains \cite{bertsch09}. In real applications, the bubbles in microchannel flow boiling often exist as bubble trains, not isolated bubbles. The interaction between the bubbles in bubble trains directly affects the heat transfer. More importantly, in real applications, the continuous operation of microchannel flow boiling leads to a stabilized condition, i.e., the process has evolved with sufficient time that the effect of the initial condition is negligible (note that it is not a static state because of the periodic pass-through of the bubbles). Such stabilized conditions cannot be reproduced by simulating single isolated bubbles or a few bubbles because the liquid temperature surrounding the bubble is affected by the heat transfer of preceding bubbles. Understanding the bubble trains in a stabilized state is important because it provides a good representation of how microchannel flow boiling functions under continuous operation conditions. The study can underscore the remarkable effect of the bubble trains on the flow dynamics and thermal performance of microchannel heat sinks.

In this study, to achieve a continuous bubble train in the microchannel, the flow-focusing technique is adopted. Flow-focusing in microchannels is a simple hydrodynamic approach used to generate monodisperse droplets or bubbles by intersecting two immiscible fluids \cite{zhu17}. The procedure of this technique is the focusing of the vapor phase (dispersed phase) as a stream at the microchannel center surrounded by the liquid phase (continuous phase) to cause the bubble breakup \cite{baroud10}. The flow-focusing geometry allows us to fine-tune the bubble size and frequency in bubble trains. The dynamics of vapor bubble trains in flow boiling microchannels is investigated numerically. The effects of the initial vapor-liquid volume ratio, wall heat flux, latent heat, and capillary number on the dynamics of the bubble trains and the heat transfer are investigated.

\section{Numerical method}\label{sec:2}
\subsection{Numerical model}\label{sec:2.1}
The open-source platform OpenFOAM is adopted to simulate bubble trains in microchannel flow boiling. The solver multiRegionPhaseChangeFlow developed by Scheufler and Roenby \cite{scheufler23} is used, and is introduced briefly here. The mass, momentum, energy, and phase fraction equations are solved. The equations of mass and momentum conservations are
\begin{equation}\label{eq:01}
  \frac{\partial (\rho \text{)}}{\partial t}+\nabla \cdot (\rho \mathbf{u})=0,
\end{equation}
\begin{equation}\label{eq:02}
  \frac{\partial (\rho \mathbf{u})}{\partial t}+\nabla \cdot (\rho \mathbf{uu})=-\nabla p-\nabla \cdot \{{{\mu }_{\text{eff}}}[\nabla \mathbf{u}+{(\nabla \mathbf{u})}^{T}]\}+\rho g+{{\mathbf{f}}_{\sigma }},
\end{equation}
where ${{\mu }_{\text{eff}}}$ and ${{\mathbf{f}}_{\sigma }}$ are the fluid viscosity and the surface tension force, respectively. The energy conservation equation adopts a two-field approach in terms of the temperature,
\begin{equation}\label{eq:03}
  \frac{\partial (\rho {{c}_{p}}T)}{\partial t}+\nabla \cdot (\rho {{c}_{p}}\mathbf{u}T)=\nabla \cdot (k\nabla T)-{{\dot{q}}_{pc}},
\end{equation}
where $k$ and ${{c}_{p}}$ are the thermal conductivity and the specific heat capacity, and ${{\dot{q}}_{pc}}$ is for energy changes caused by phase change.

We use the volume of fluid (VOF) method to solve the phase fraction equation and capture the liquid-vapor interface, which is expressed as
\begin{equation}\label{eq:04}
  \frac{\partial \alpha }{\partial t}+\nabla \cdot (\mathbf{u}\alpha )={{\dot{\alpha }}_{pc}},
\end{equation}
where $\alpha$ is the liquid volume fraction, and ${{\dot{\alpha }}_{pc}}$ is the explicit source term to account for phase change. 

The thermophysical parameters of the fluid are calculated from the volume fraction, 
\begin{equation}\label{eq:05}
  \varphi =(1-\alpha ){{\varphi }_{v}}+\alpha {{\varphi }_{l}},
\end{equation}
where $\varphi $ represents any fluid properties, including viscosity, density, and thermal conductivity.

The surface tension force at the liquid-vapor interface ${{\mathbf{f}}_{\sigma }}$ in Eq.\ (\ref{eq:02}) is \cite{Brackbill1992ModelingSurfaceTension}
\begin{equation}\label{eq:06}
  {{\mathbf{f}}_{\sigma }}=\sigma \kappa \mathbf{n}\left| \nabla \alpha  \right|\frac{2\rho }{{{\rho }_{v}}+{{\rho }_{l}}},
\end{equation}
where $\kappa $ and $\mathbf{n}$ are the curvature and the unit normal vector of the liquid-vapor interface,
\begin{equation}\label{eq:07}
  \kappa =-\nabla \cdot \mathbf{n},
\end{equation}
\begin{equation}\label{eq:08}
  \mathbf{n}=\frac{\nabla \alpha }{\left| \nabla \alpha  \right|}.
\end{equation}

The Hardt and Wandra model \cite{hardt08} is used to calculate the mass transfer due to phase change at the liquid-vapor interface. More details of the phase-change model can be found in Refs.\ \cite{hardt08, scheufler23}. The phase change adopted to calculate the energy source term in Eq.\ (\ref{eq:03}) is a gradient-based model, which is computed implicitly to improve the solver's stability and expressed as
\begin{equation}\label{eq:09a}
\dot{q}_{pc}=q_{pc}^{l}+q_{pc}^{v}={{k}^{l}}\nabla {{T}^{l}}\cdot {{\mathbf{\hat{n}}}_{s}} \frac{A_{\text{int}}}{V} +{{k}^{v}}\nabla {{T}^{v}}\cdot (-{\mathbf{\hat{n}}_{s}})\frac{A_{\text{int}}}{V}
\end{equation}
where ${\mathbf{\hat{n}}_{s}}$ is the interface normal, $A_\text{int}$ is the interface surface area in the
cell, and $V$ is the cell volume. The mass transfer in Eq.\ (\ref{eq:04}) is evaluated as 
\begin{equation}\label{eq:10a}
{{\dot{\alpha }}_{pc}}=\frac{{\dot{\rho }}}{{{\rho }_{l}}}
\end{equation}
\begin{equation}\label{eq:11a}
\dot{\rho }=\frac{{\dot{q}_{pc}}}{{{h}_{lv}}}
\end{equation}
where $\dot{\rho}$ is the volume-specific mass flux at the interface and $V$ is the cell volume.

\subsection{Simulation setup}\label{sec:2.2}
\begin{figure}
  \centering
  \includegraphics[scale=0.7]{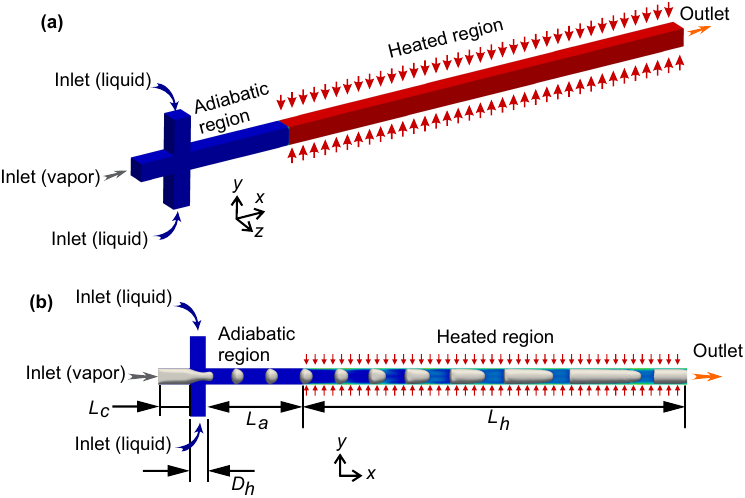}
  \caption{Simulation setup for the bubble train in microchannel flow boiling. (a) Overview of the simulation domain. (b) Dimensions of the microchannel and the boundary conditions.}\label{fig:01}
\end{figure}

Microchannel flow boiling at the slug flow regime in saturated condition consists of a continuous bubble train in the microchannel, where the main heat transfer mechanisms are the vaporization at the liquid film separating the channel wall and the bubble and the convection of the vapor bubbles and liquid slugs. To form bubble trains in microchannels, a flow-focusing geometry is used in the upstream of the microchannel (see Figure \ref{fig:01}). Although we do not include bubble nucleation in the simulation, neglecting bubble nucleation is generally acceptable in analyzing the bubble dynamics of flow boiling heat transfer \cite{Mukherjee2005BubbleGrowth, mukherjee11, Ferrari2018SlugFlowBoiling, luo20, Liu2017BubbleTrainMicrochannel, Luo2017NumericalBubbleGrowth}. In addition, the flow-focusing geometry allows us to control the size, frequency, and separation of the bubbles in the microchannel. The channel has a uniform square cross-section of $50 \times 50$ $\upmu$m$^2$, i.e., the hydraulic diameter is $D_h = 50$ $\upmu$m. Due to the very small hydraulic diameter, the influence of gravity becomes less significant and can be neglected because the Bond number, $Bo = \rho g D_h^2 /\sigma$, is on the order of $10^{-3}$. The left inlet is for the introduction of the vapor phase, and the top and bottom inlets are for the introduction of the liquid phase, as shown in Figure \ref{fig:01}(b). The lengths of the three inlets are all $L_c = 2D_h$. After the flow-focusing junction, the fluids enter a straight section of length $L = 30D_h$. The straight part of the microchannel includes an adiabatic section of length $L_a = 6D_h$ for the development of the flow, and a heated section of length $L_h = 24D_h$ with a uniform wall heat flux on the wall.

The flow-focusing technique in this setup involves the convergence of the two immiscible fluids in a way that the continuous phase (liquid phase) squeezes or focuses the dispersed phase (vapor phase) into the microchannel to form bubbles \cite{Anna2016DropletsBubbles, Anna2003FormationDispersions, Moragues2023DropletMicrofluidics}. At the inlets of the cross-flow microchannel, the inlet of the liquid phase intersects with the inlet of the vapor phase which are at a right angle to each other to create the flow-focusing zone. The generation of the bubble trains is tuned by adjusting the flow of the gas phase and the liquid phase at the flow-focusing geometry. A constant liquid inlet velocity is adopted at the cross-junction ($U_l = 0.31$ m/s) with a Reynolds number of $Re = {{{\rho }_{l}}{{U}_{l}}{{D}_{h}}}/{{{\mu }_{l}}} = 100$. The vapor inlet velocity is varied from $0.4U_l$ to $0.6U_l$, for the formation of bubbles at different sizes and frequencies. These correspond to the inlet vapor-liquid volume ratios of 0.2 to 0.3 (note that there are two liquid inlets). 

A quarter of the computational domain is simulated to save computational cost, and the symmetry condition is applied to the middle cross-sections of both the $y$ and $z$ directions. The working fluid is R134a with the properties listed in Table $\ref{tab:1}$. The inlet temperature is set to the fluid saturation temperature $T = T_\text{sat}$ for both the liquid and vapor phases. Unless otherwise specified, the wall heat flux in the heated region is $q_w = 150$ kW/m$^2$. The initial temperature of the whole domain is $T_\text{sat}$, and the domain temperature changes with time and gradually reaches a stabilized state, in which the process is time periodic because bubbles are formed periodically. Hence, in this study, we focus on the flow dynamics and heat transfer of the bubble train in the stabilized state. 

\begin{table}[]
\centering
\caption{Thermophysical properties of the working fluid R134a.}
\label{tab:1}
\begin{tabular}{lcc}
\hline
Properties                                       & R134a (liquid) & R134a (vapor) \\ \hline
Density, $\rho$ [kg/m$^3$]                       & 1187.5         & 37.535        \\
Dynamic viscosity, $\mu$ [mPa s]           & 0.1846         & 0.01238       \\
Thermal conductivity, $k$ [mW/(m K)]       & 80.27          & 15.01         \\
Specific heat capacity, $c_p$   [J/(kg K)] & 1446           & 1065          \\
Saturation temperature, $T_\text{sat}$   [K]          & \multicolumn{2}{c}{303.15}     \\
Surface   tension, $\sigma$ [mN/m]               & \multicolumn{2}{c}{7.56}       \\
Latent heat, $h_{lv}$ [kJ/kg]                    & \multicolumn{2}{c}{173.1}      \\ \hline
\end{tabular}
\end{table}

To characterize the convection at the heated wall, the average Nusselt number $\overline{\Nu}$ is computed from the average convective heat transfer coefficient $\overline{h}$,
\begin{equation}\label{eq:09}
  \overline{\Nu}=\frac{\bar{h}{{D}_{h}}}{{{k}_{l}}},
\end{equation}
\begin{equation}\label{eq:10}
  \bar{h}=\frac{1}{A}\int_{0}^{A}{hdA},
\end{equation}
\begin{equation}\label{eq:11}
  h=\frac{q''}{{{T}_{w}}-{{T}_\text{sat}}},
\end{equation}
where $T_w$ and $q''$ are the local temperature and local heat flux at the heated wall.

\subsection{Mesh independence study and validation}\label{sec:2.3}
\begin{figure}
  \centering
  \includegraphics[scale=0.2]{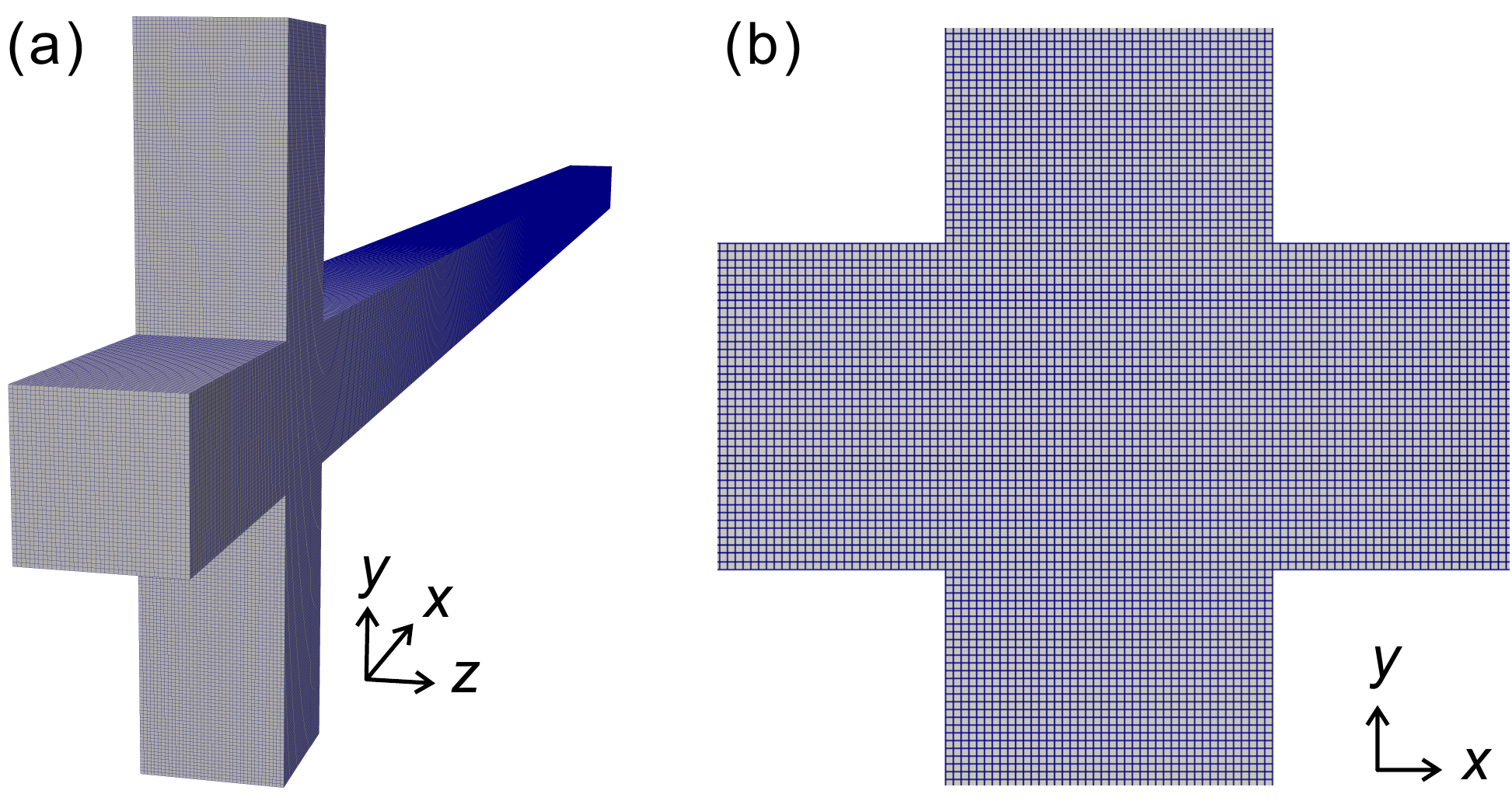}
  \caption{(a) Mesh for the 3D simulation domain. (b) Mesh at the flow-focusing junction.}\label{fig:02}
\end{figure}

\begin{figure}
  \centering
  \includegraphics[scale=0.27]{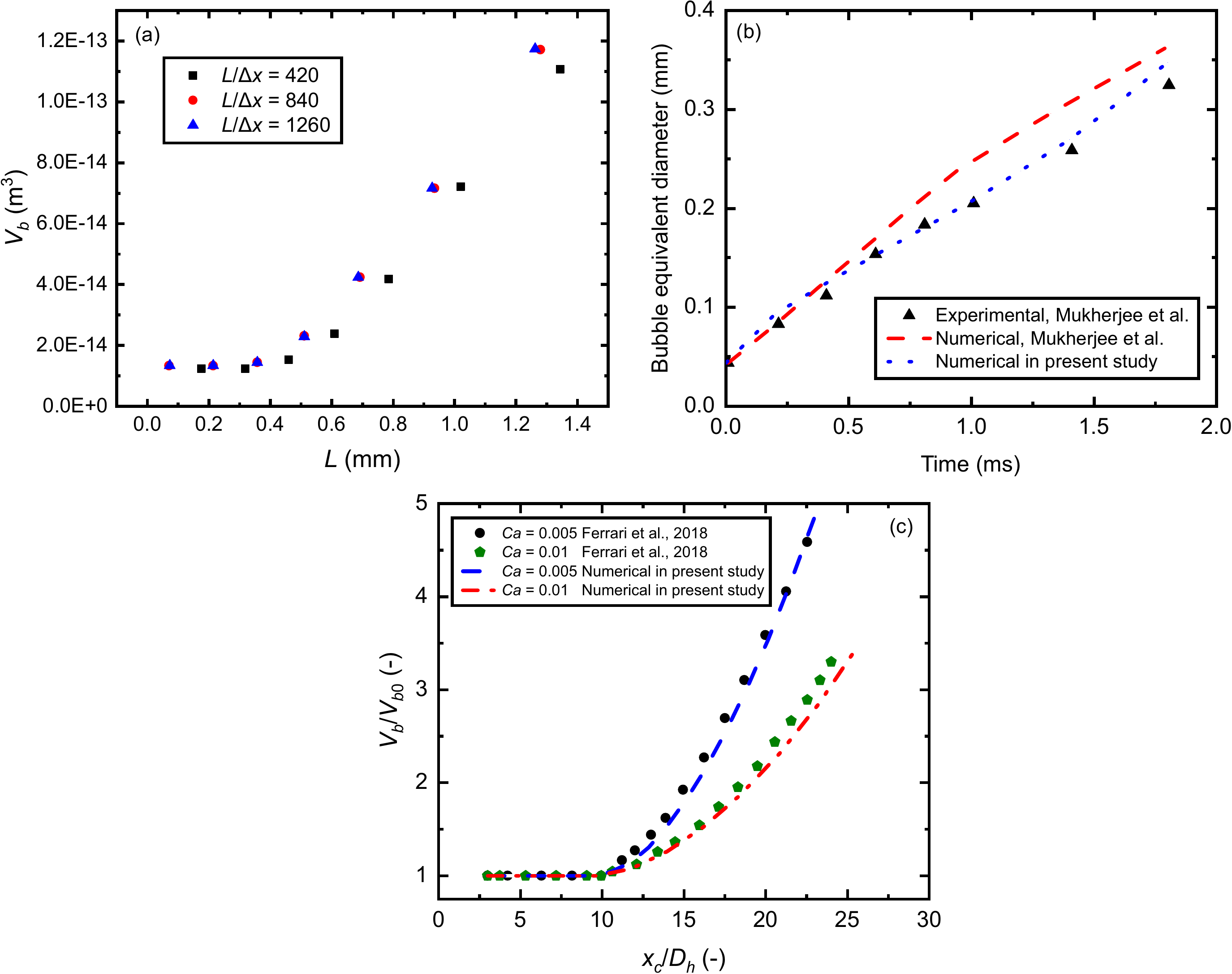}
  \caption{(a) Bubble volume $V_b$ versus bubble center position for the bubble train in the grid independence study. The vapor-liquid volume ratio is 0.25 and the time is $t = 10$ ms. (b) Time variation of the bubble equivalent diameter compared with the results obtained by Mukherjee et al.\ \cite{mukherjee11}. (c) Dimensionless bubble volume $V_b/V_{b0}$ versus bubble center position $x_c/D_h$ compared with the results obtained by Ferrari et al.\ \cite{Ferrari2018SlugFlowBoiling}.}\label{fig:03}
\end{figure}

The simulation domain is uniformly meshed with a highly refined grid with structured orthogonal hexahedral cells as shown in Figure \ref{fig:02}. To balance the computational cost and accuracy, a mesh independence study is conducted to determine if the mesh resolution can adequately capture the bubble dynamics. Three mesh resolutions are tested, namely $L/\Delta x =$ 420, 840, and 1260, where $L/\Delta x$ is the mesh size in $x$ direction. The results of $L/\Delta x =$ 840 and 1260 have a minute difference as shown in Figure \ref{fig:03}(a), thus the mesh resolution of $L/\Delta x =$ 840 is adopted for further simulations. The simulation is also validated against the experimental results of flowing bubble growth in a microchannel by Mukherjee et al.\ \cite{mukherjee11} and the results obtained by Ferrari et al.\ \cite{Ferrari2018SlugFlowBoiling}, which proves the accuracy of the numerical results, see Figure \ref{fig:03}(b,c). While the validation was conducted for a single bubble, the underlying physics governing the bubble growth and behavior are related to multiple bubbles (bubble trains). The numerical model incorporates the same principles of flow dynamics and heat transfer that are applicable to both single and bubble train scenarios. Therefore, the validated single-bubble results provide confidence in the model's ability to simulate bubble trains under similar conditions. In the experiment by Mukherjee et al.\ \cite{mukherjee11}, saturated water at the temperature of 373K and atmospheric pressure entered with 0.146 m/s uniform velocity into a square microchannel with a hydraulic diameter of $D_h = 229 \upmu$m and $5D_h$ length. The channel is heated with a constant temperature of 375.1K imposed on three sides with the top fourth wall set as adiabatic. The walls are set as no-slip boundary conditions and simulated with our numerical model.
    
\section{Results and discussion}\label{sec:3}
\subsection{Effect of initial vapor-liquid volume ratio (${\VR}$)}\label{sec:3.1}
\begin{figure}
  \centering
  \includegraphics[width=\columnwidth]{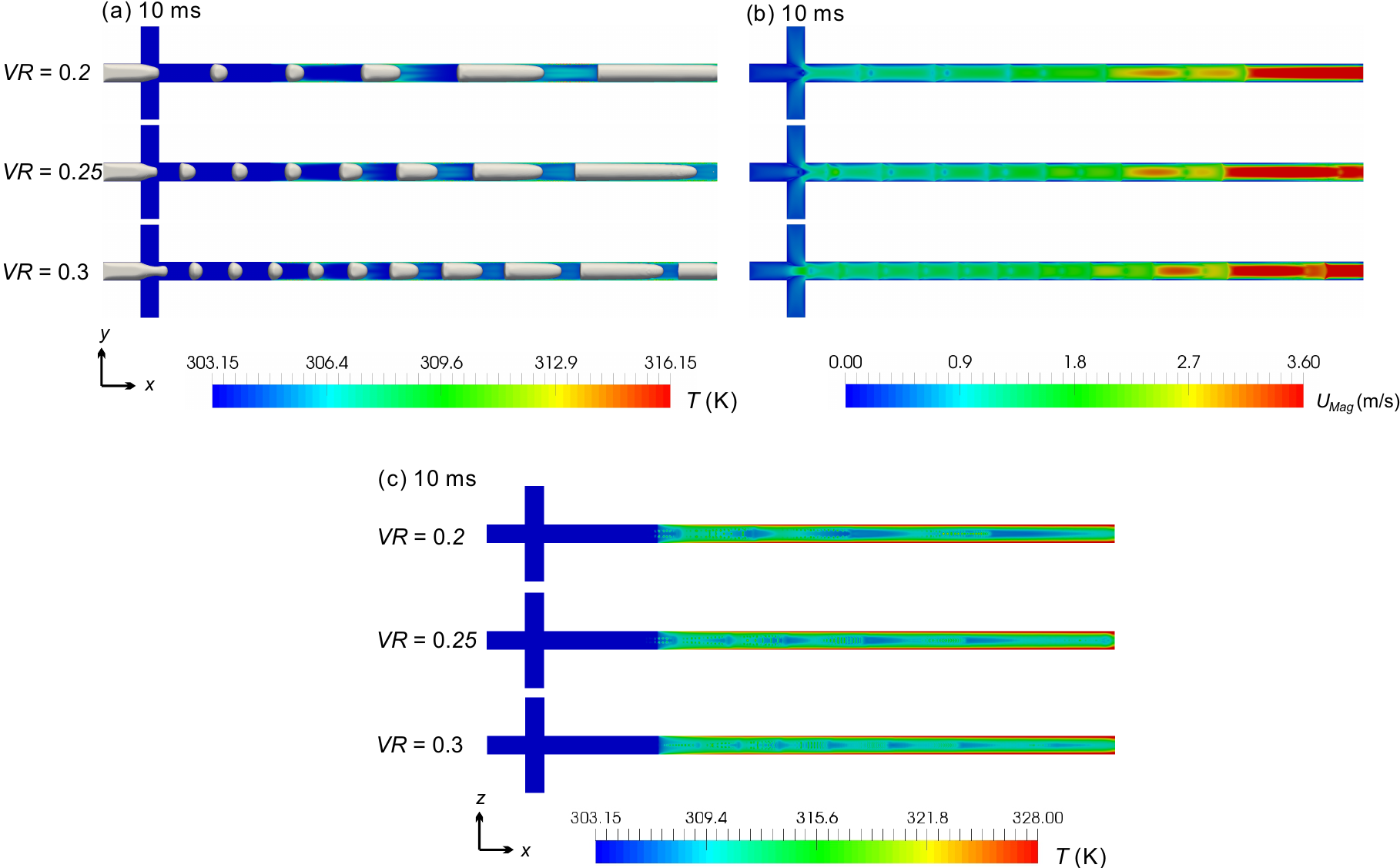}
  \caption{Bubble train in microchannel flow boiling for different volume ratios at a typical instant of the stabilized state: (a) bubble shapes and temperature fields in the middle cross-section in the $z$-direction; (b) velocity field at middle cross-section in the $z$-direction; (c) bottom wall temperature illustrating the temperature distribution of the heated wall for different volume ratios in the stabilized state.}\label{fig:04}
\end{figure}

\begin{figure}
  \centering
  \includegraphics[scale=0.27]{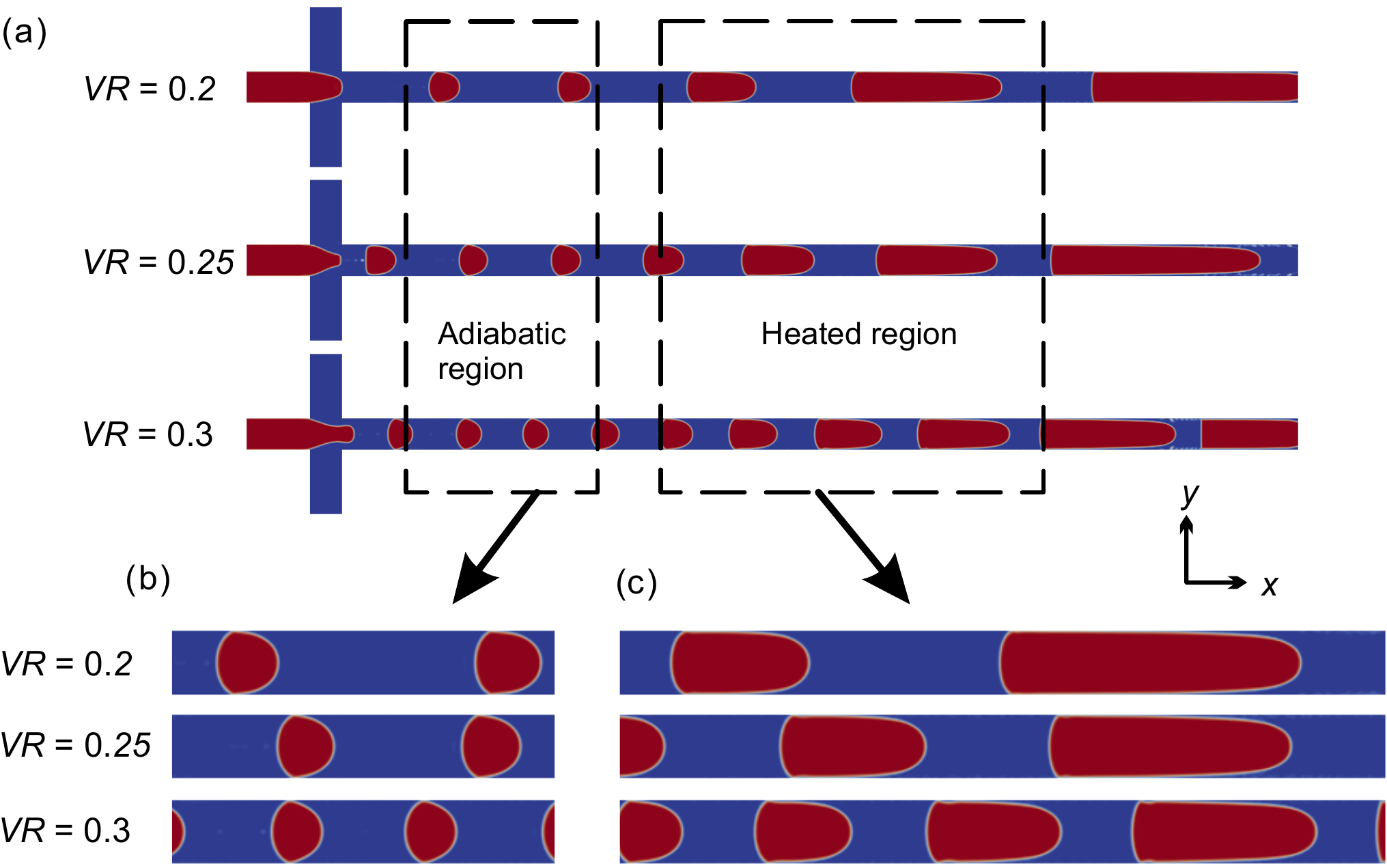}
  \caption{Bubble train phase fraction field in microchannel flow boiling for different velocity ratios at a typical instant of the stabilized state in the middle cross-section in the $z$-direction: (a) overview of the entire channel, (b) zoom-in of the adiabatic region, (c) zoom-in of the heated region.}\label{fig:05}
\end{figure}

The volume fraction of the vapor phase is an important factor for the heat transfer in the downstream of the microchannel. The effect of the initial vapor-liquid volume ratio (${\VR}$) on the vapor bubble train is studied by varying the inlet vapor flow rate while keeping the inlet liquid flow rate constant (Figure \ref{fig:04}). The process of bubble formation at the flow-focusing geometry involves the squeezing of the vapor phase by the two liquid streams. After the formation of monodispersed bubbles, the bubble shape evolves in the adiabatic region. With increasing ${\VR}$, the bubble size decreases and the bubble spacing in the microchannel decreases, as shown in Figure \ref{fig:04}(a). The flow boiling process starts when the bubble train enters the heated section. The bubble size increases quickly and the bubble shapes change while moving downstream due to vaporization, see Figure \ref{fig:04}(a). For different ${\VR}$, the bubble size at a low ${\VR}$ is generally much larger than that of a high ${\VR}$. This is because the larger initial bubble size increases the contact of the bubble interface with the superheated thermal boundary layer, leading to faster bubble growth, as shown in Figure \ref{fig:05}(b). 

The flow fields of the bubble trains with different ${\VR}$ are shown in Figure \ref{fig:04}(b). The velocity in the downstream is much larger than that in the upstream. The main reason is the expansion of the bubbles, which pushes the bubbles and the liquid slugs in the downstream to achieve higher speeds. The expansion rates of the bubbles are different along the channels, and the bubbles in the downstream have higher expansion rates. Hence, the velocity gradient along in the downstream is much higher than that in the upstream. Particularly, the velocity variation near the tail of the bubbles in the downstream is very large, indicating the velocity in the vapor bubble is much larger than that in the liquid slug. This could be attributed to the rapid vaporization of fluid at the tails of the downstream bubbles. In these regions, the liquid film beneath the bubble is very thin, as shown in Figure \ref{fig:05}(c). As the liquid film becomes thinner, the heat transfer is enhanced because the local thermal resistance is reduced leading to the larger bubble growth. In addition, the bubble produces strong perturbation to the flow. Hence, the heat transfer here is very effective and the fluid vaporization is very fast, inducing the acceleration of the bubble. The velocity fields produced by the bubble train have a strong impact on the temperature field of the heated wall, as shown in Figure \ref{fig:04}(c). An important feature of the temperature distribution on the heated wall is that the temperature near the two side walls is much higher than the center. This is because of the thickness of the liquid film beneath the bubble along the cross-section of the microchannel. In the corner of the square cross-section of the channel, without the perturbation by the bubble, the flow speed is very slow and the thermal boundary layer is very thick. Hence, the convective heat transfer at the corner is very ineffective. When the bubbles flow in the microchannel, the bubbles significantly perturb the flow fields, which then enhance the convection from the wall to the fluid, reducing the wall temperature. The wall temperature distribution along the channel direction also reflects the periodic nature of bubble trains. As bubbles pass through the channel, they induce alternating regions of low and high temperatures. The temperature is lower beneath the bubble than the regions between the bubbles, where the liquid slug is present. This is because of the heat absorption by the vaporization of the fluid, which reduces the local temperature of the fluid and also the local temperature of the wall.
 
\begin{figure}
  \centering
  \includegraphics[scale=0.3]{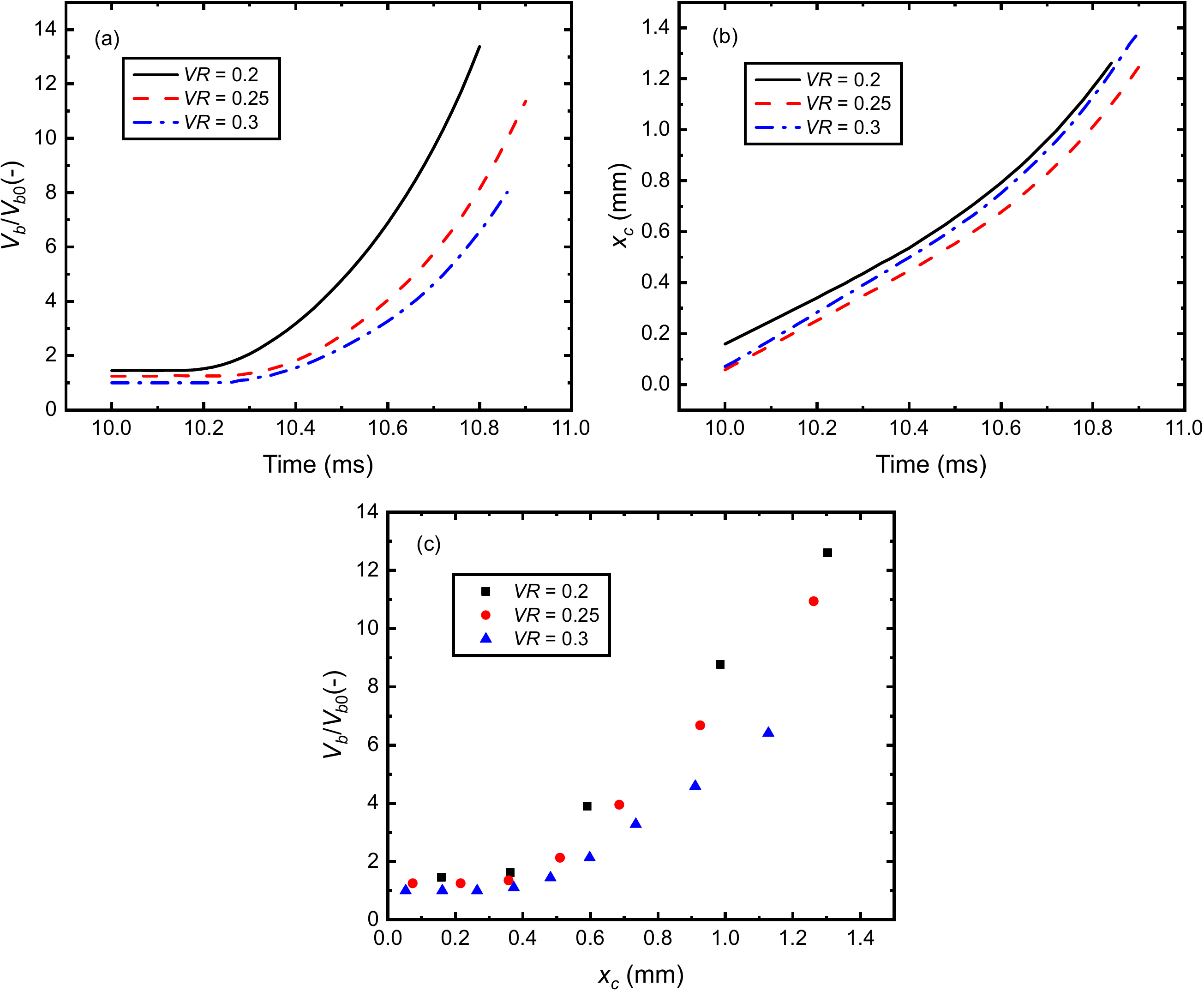}
  \caption{Effects of initial vapor-liquid flow rate ratio on the volume and position of the bubbles in the stabilized state: (a) Time variation of the dimensionless bubble volume $V_b/V_{b0}$ of a single bubble; (b) Time variation of the dimensionless bubble position $x_c/D_h$ of a single bubble; (c) Dimensionless bubble volume $V_b/V_{b0}$ versus bubble center position at a typical instant.}\label{fig:06}
\end{figure}

To quantitatively discuss the growth of the bubbles in the microchannel at different ${\VR}$, the time variation of the dimensionless bubble volume $V_b/V_{b0}$ of a single bubble in the bubble train in the stabilized state for different volume ratios is plotted in Figure \ref{fig:06}(a). Initially, the bubble produced at a lower ${\VR}$ is larger, and its size grows quicker than that at a higher ${\VR}$. This is because of the faster growth of larger bubbles, which have closer contact with the superheated thermal boundary layer, as discussed in Figure \ref{fig:04}. The dimensionless bubble volumes of the bubble train at a typical instant are shown in Figure \ref{fig:06}(c). Each symbol in the plot represents a bubble in the bubble train, and the size variation of the bubbles is consistent with the time variation of the bubbles. The time variations of the bubble positions for typical bubbles are shown in Figure \ref{fig:06}(b). We can see that the curves become steeper when the bubbles flow further to the downstream, indicating the bubble speeds are becoming faster. This is mainly because of the acceleration of the bubbles caused by liquid vaporization and bubble expansion. If compare the three curves of different values of ${\VR}$, we can see that their initial positions are different because the bubbles are generated at the upstream at different frequencies. Hence, we can compare their slopes, which indicate the speeds of the bubbles. As the ${\VR}$ increases, we can see the slope increases, indicating the bubbles are moving faster. This is because of the decrease in the bubble size, which increases the liquid film thickness beneath the bubble and reduces the friction of the bubble movement.

\begin{figure}
  \centering
  \includegraphics[scale=0.5]{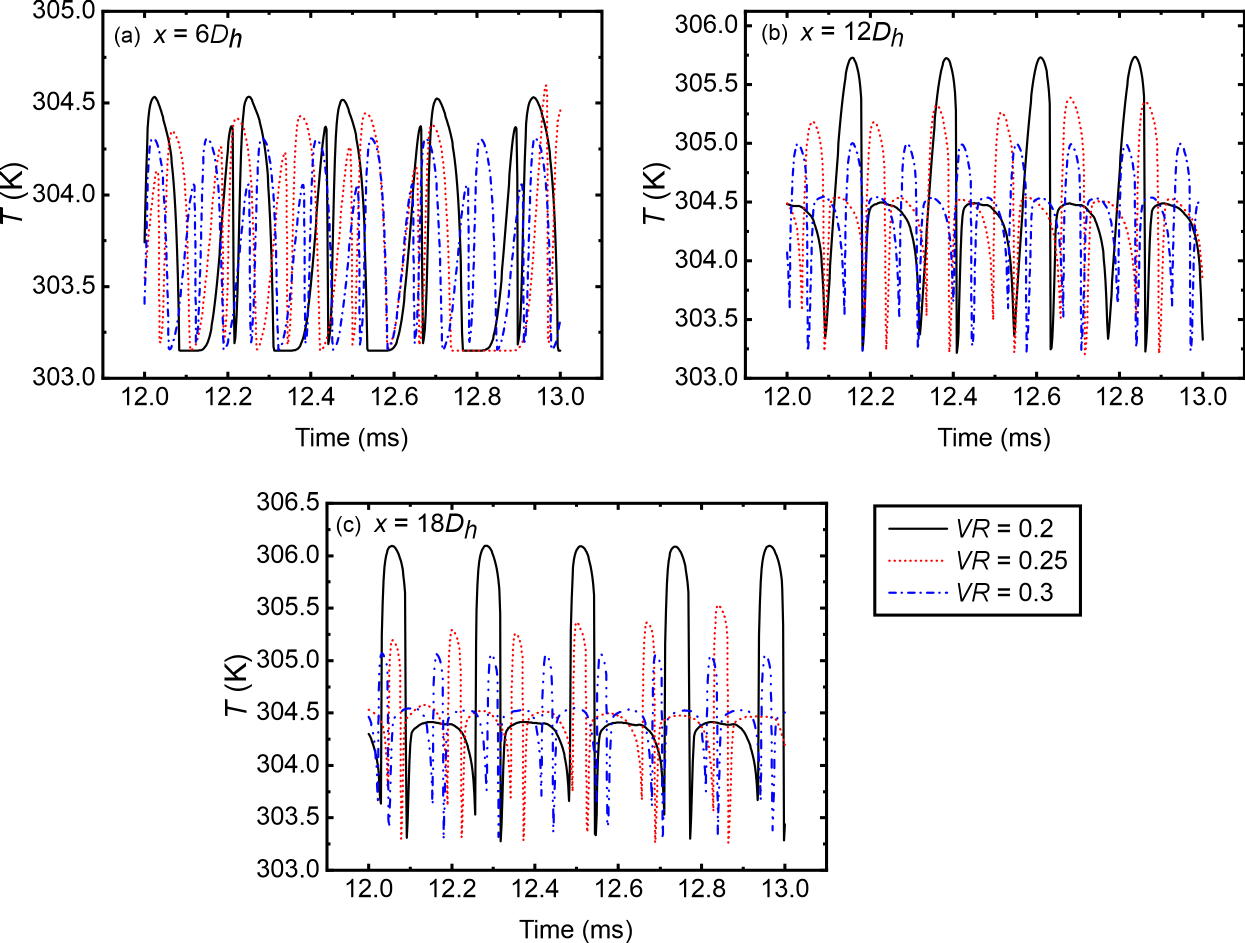}
  \caption{Time variation of the wall temperature at typical axial locations at the mid-point of the heated wall for different ${\VR}$: (a) $x = 6D_h$, (b) $x = 12D_h$, and (c) $x = 18D_h$.}\label{fig:07}
\end{figure}

To examine the effect of the ${\VR}$ on the heat transfer, the time variation of the wall temperatures at several typical axial locations of the microchannel is extracted and shown in Figure \ref{fig:07}. We can clearly see that the wall temperature fluctuates periodically, which is due to the periodic pass-through of the bubbles. With the increase in the ${\VR}$, the magnitude of the temperature fluctuation decreases. This could be attributed to the larger number of bubbles and smaller bubble sizes, which results in a shorter time for temperature increase, hence weaker temperature fluctuation. In addition, the temperature fluctuation in the downstream is much higher than that in the upstream as the bubble size increases along the channel. 

\begin{figure}
  \centering
  \includegraphics[scale=0.3]{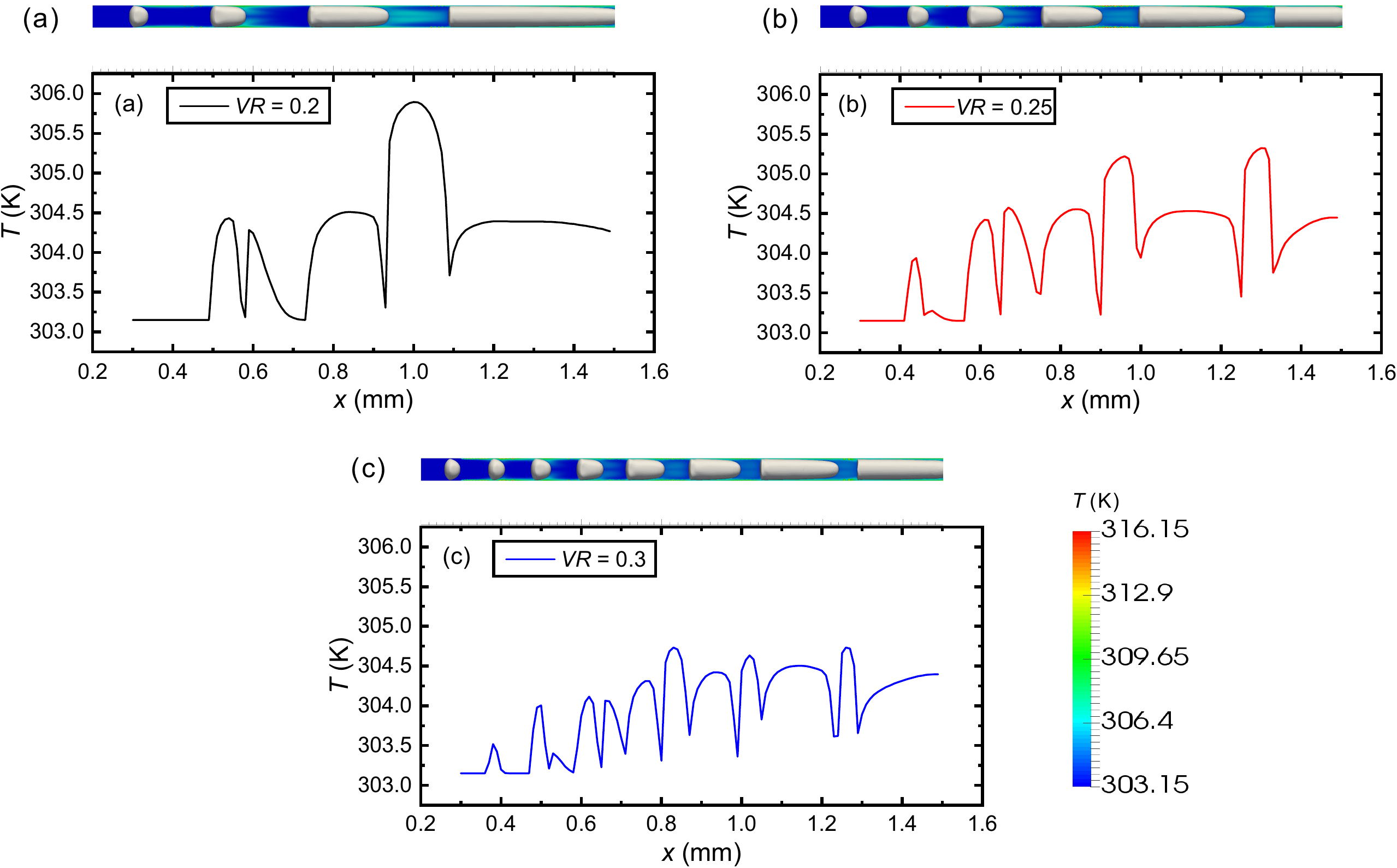}
  \caption{Temperature distribution along the center line of the heated bottom wall ($z = 0$) and the corresponding temperature field in the middle cross-section in the $z$-direction for different ${\VR}$ at t = 12.2 ms: (a) ${\VR}$ = 0.2, (b) ${\VR}$ = 0.25, (c) ${\VR}$ = 0.3.}\label{fig:08}
\end{figure}

To understand the temperature fluctuation along the channel for different ${\VR}$, the temperature distribution along the center line of the heated bottom wall is plotted for a typical instant in Figure \ref{fig:08}. In the upstream, at the adiabatic region, the temperature is uniform at saturation temperature. In addition, at the upstream of the heated section, the wall temperature at the front and the tail of the bubble are low, and the wall temperature beneath the bubbles and the slugs is higher. This could be attributed to the rapid vaporization at the front and the tails of the bubbles, which absorbs latent heat and reduces the local temperature. In the downstream, the wall temperatures beneath the bubbles and beneath the liquid slugs show different trends. The wall temperature beneath the bubbles is at about 304.5 K and does not change much. This is because the bubble temperature is very close to the boiling point and the heat transfer resistance is mainly the liquid. In contrast, the wall temperature beneath the slugs increases along the channel. This is because the liquid in the slug is superheated, the heat accumulates in the slug, and the temperature increases continuously. With the increase in the ${\VR}$, we can see that the wall temperature beneath the bubbles does not change much, but the wall temperature beneath the slugs is significantly affected, i.e., the magnitude of temperature variation becomes smaller, and the frequency of the temperature variation becomes higher. This is mainly because the higher bubble frequency at higher ${\VR}$ allows less time for the heat transfer from the wall to the slug, resulting in temperature fluctuation with smaller amplitudes but higher frequencies.

\begin{figure}
  \centering
  \includegraphics[scale=0.31]{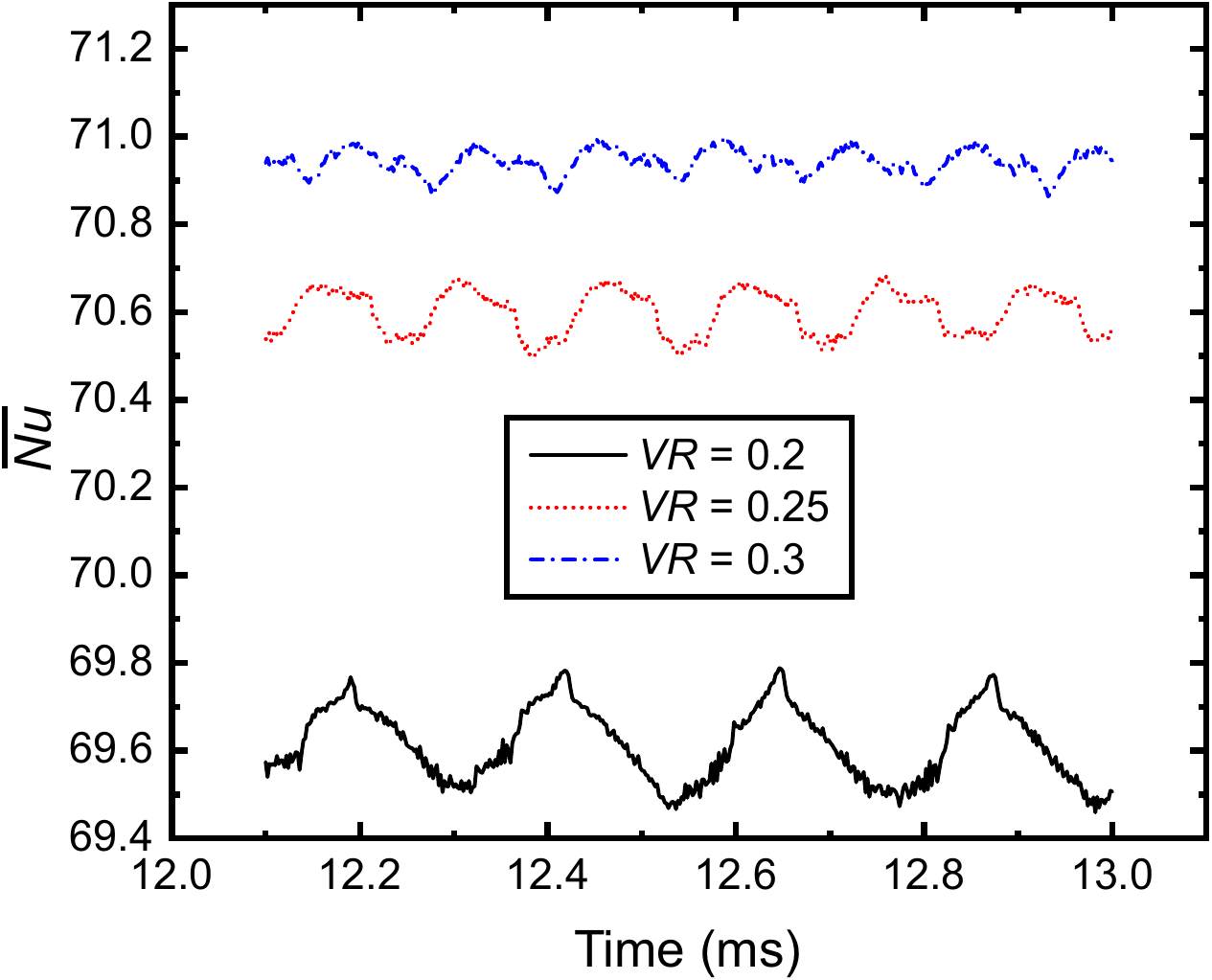}
  \caption{Average Nusselt number $\overline{\Nu}$ versus time for different ${\VR}$.}\label{fig:09}
\end{figure}

As a consequence of the temperature oscillations, the Nusselt number is remarkably affected by the bubble train. The average Nusselt number at the microchannel heated wall $\overline{\Nu}$ is plotted against time in Figure \ref{fig:09}. The Nusselt number exhibits fluctuations against time. The fluctuations correspond to the periodic formation of the bubbles in microchannels. As the ${\VR}$ increases, the Nusselt number increases with higher-frequency fluctuations as shown in Figure \ref{fig:09}. This is because increasing ${\VR}$ enhances the bubble frequency and with more bubbles present within the microchannel, and the heat transfer is enhanced.

\subsection{Effect of wall heat flux}\label{sec:3.2}
\begin{figure}
  \centering
  \includegraphics[width=\columnwidth]{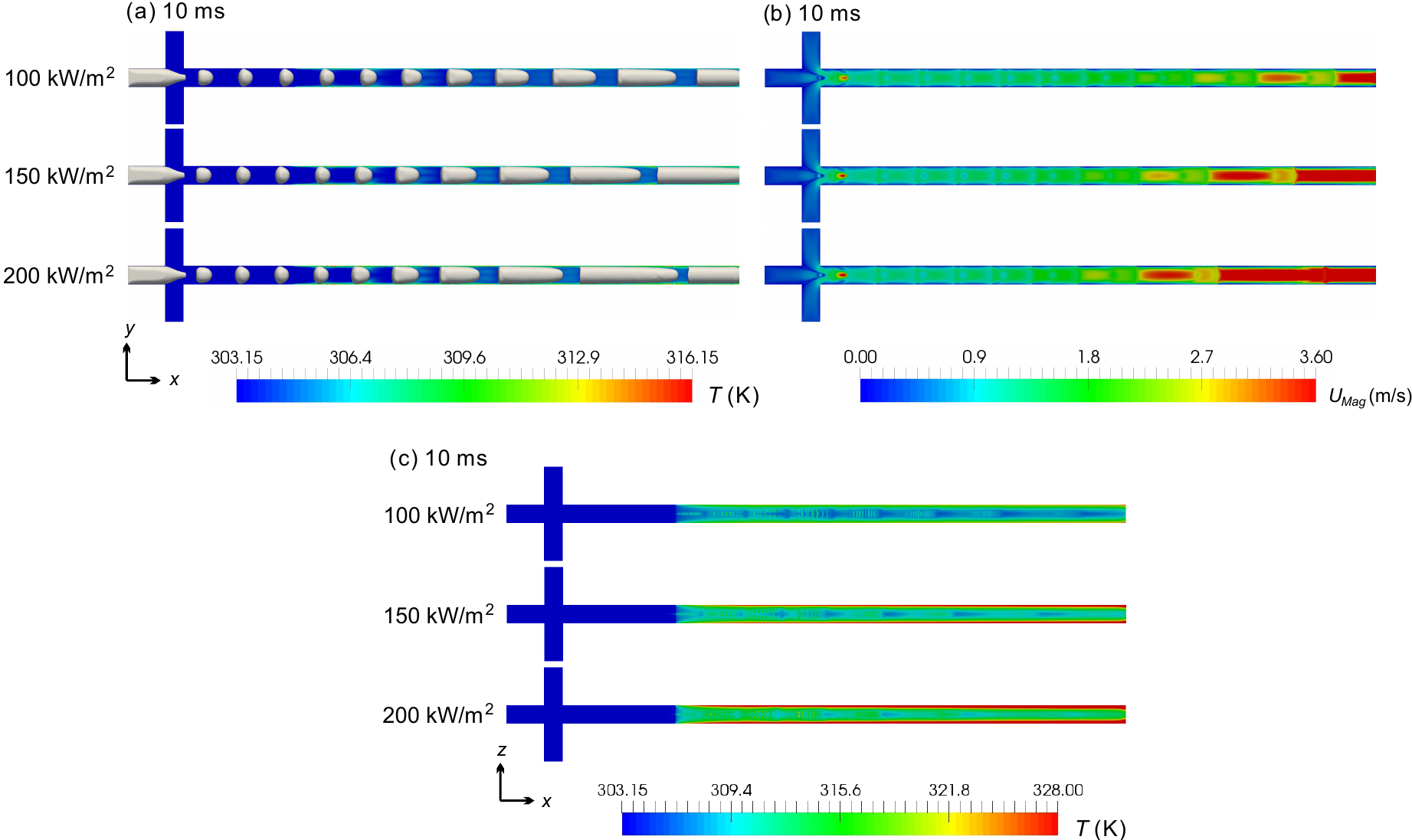}
  \caption{Bubble train in microchannel flow boiling for different heat fluxes at a typical instant of the stabilized state: (a) bubble shapes and temperature fields in the middle cross-section in the $z$-direction; (b) velocity field in the middle cross-section in the $z$-direction; (c) bottom wall temperature illustrating the temperature distribution of the heated wall for different heat fluxes in the stabilized state.}\label{fig:10}
\end{figure}

The wall heat flux $q_w$ is a key parameter for the boiling heat transfer. Figure \ref{fig:10}(a, b) shows the temperature and the flow fields for different heat fluxes $q_w$ at a typical instant in the stabilized state. The bubbles have the same initial sizes when flowing downstream at the adiabatic region. However, the growth of the bubble train varies for different $q_w$. With increasing $q_w$, the bubble train growth within the channel increases due to the thicker thermal boundary layer in the channel, subsequently increasing the vaporization rate at the vapor-liquid interface. The microchannel with a lower $q_w$ has more bubbles present in the microchannel at a typical instant because the bubble expansion is reduced compared to that with a larger $q_w$. The reduction in the bubble growth gives more space for more bubbles to be present within the channel. 

The corresponding flow fields at different $q_w$ are presented in Figure \ref{fig:10}(b). We can see that the flow in the microchannel is more significantly perturbed as $q_w$ increases. Immediately after the bubble leaves the adiabatic region, only the velocity in the very small area near the bubbles is affected. However, in the downstream, the flow velocity is significantly accelerated. As $q_w$ increases, the perturbation to the flow field increases due to increased vaporization at the heated region, and subsequently enhanced the heat transfer by the convection effect. With increasing $q_w$, the temperature of the heated wall increases significantly, as shown in Figure \ref{fig:10}(c), particularly near the two side walls. In the liquid film near the two side walls, the flow is very weak, and the heat transfer from the wall to the liquid is mainly by thermal conduction \cite{che20153d}. Hence, at higher $q_w$, a higher temperature is required to produce the local temperature gradient.

\begin{figure}
  \centering
  \includegraphics[scale=0.27]{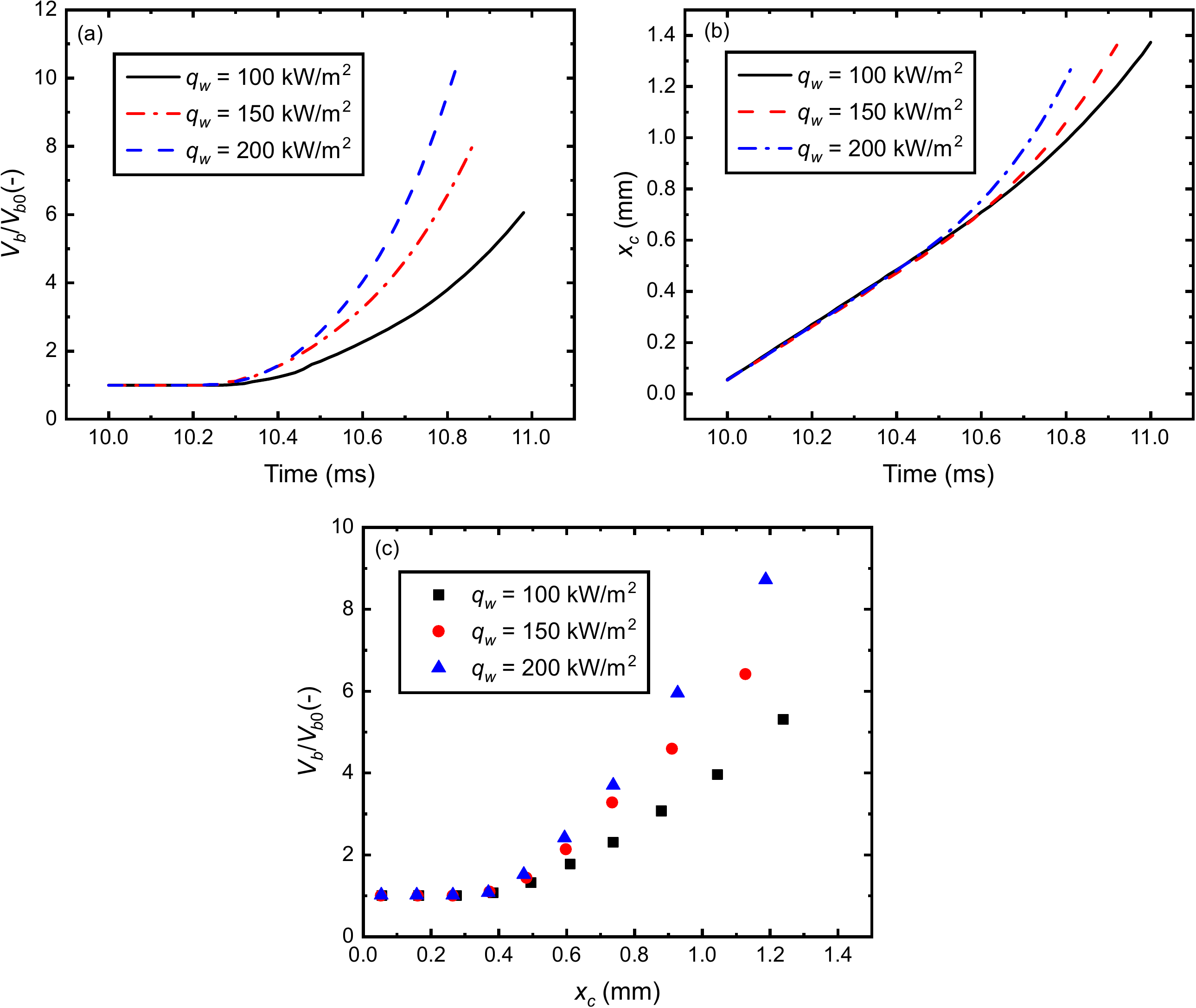}
  \caption{Effects of different heat fluxes on the volume and position of the bubbles in the stabilized state: (a) Time variation of the dimensionless bubble volume $V_b/V_{b0}$ of a single bubble; (b) Time variation of the dimensionless bubble position $x_c/D_h$ of a single bubble; (c) Dimensionless bubble volume $V_b/V_{b0}$ versus bubble center position at a typical instant.}\label{fig:11}
\end{figure}

Further, to evaluate the influence of different $q_w$ on the bubble growth, the time variation of the dimensionless bubble volume $V_b/V_{b0}$, the time variation of the center position of a bubble in the bubble train, and the dimensionless bubble volumes of the bubble train at a typical instant at stabilized state are plotted in Figure \ref{fig:11}. Before the arrival of the bubbles at the heated section, the bubble sizes remain unchanged, and then they increase along the channel as the bubbles get in contact with the superheated thermal boundary layer at the heated wall. The channel with the highest heat flux has the largest bubble growth rate because of the increased heat absorption of the bubble from the wall, as shown in Figure \ref{fig:11}(a). The variation of bubble center positions for different heat fluxes is plotted in Figure \ref{fig:11}(b). In the early stage, the bubble moves in the adiabatic region at a constant speed because there is no phase change. Then in the heated region, the bubble accelerates in the microchannel due to the expansion of the bubbles. With increasing $q_w$, the bubble acceleration becomes larger because of faster vaporization at higher $q_w$. At a typical instant, the sizes of the bubbles in the microchannels are plotted against the bubble positions, as shown in Figure \ref{fig:11}(c). We can see that the bubbles in the downstream of the channels are larger than that in the upstream. In addition, the bubbles in microchannels with higher $q_w$ are larger than those with lower $q_w$, which is consistent with the trend of bubble growth presented in Figure \ref{fig:11}(a). 

\begin{figure}
  \centering
  \includegraphics[scale=0.6]{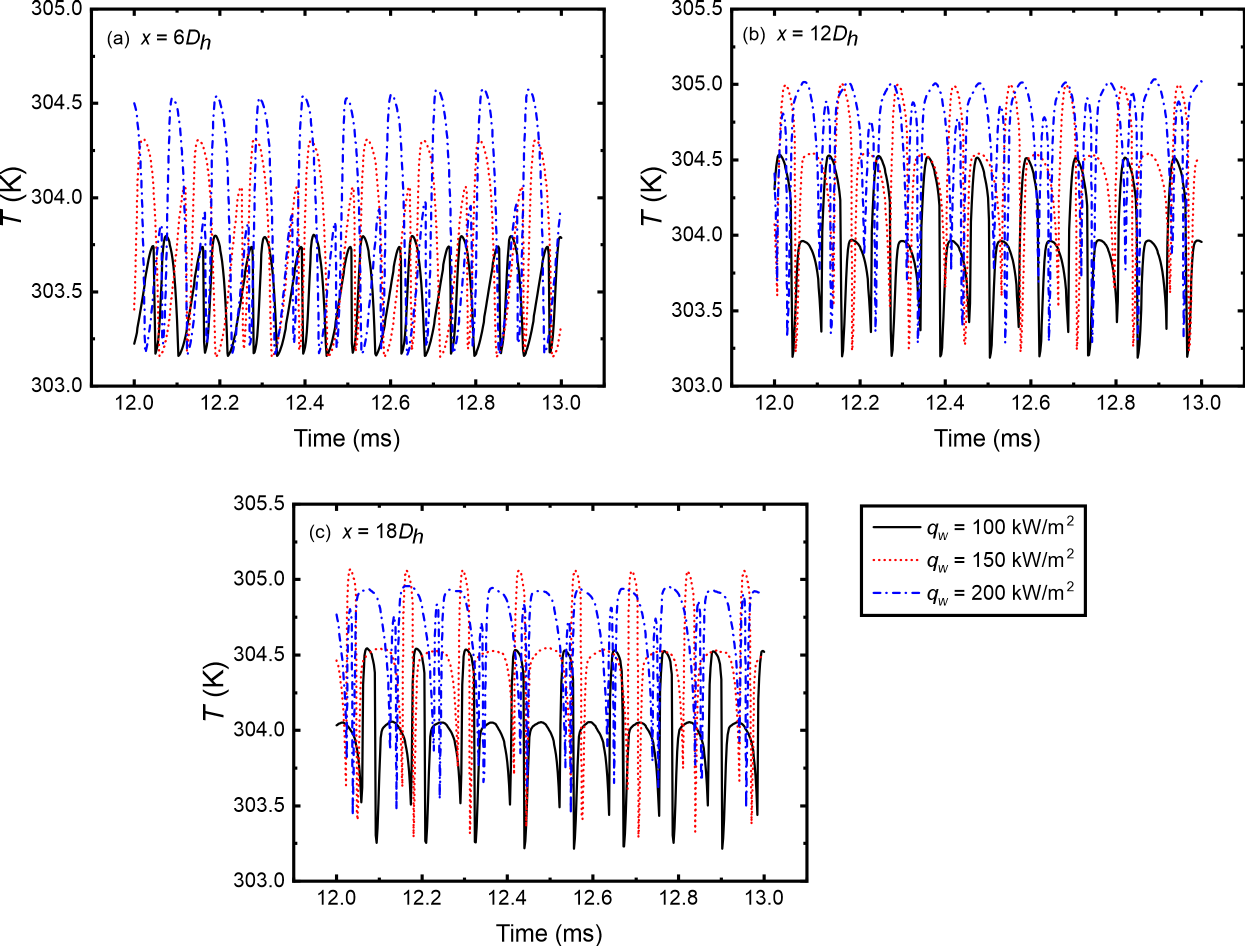}
  \caption{Time variation of the wall temperature at typical axial location at the mid-point of the heated wall for different heat fluxes: (a) $x = 6D_h$, (b) $x = 12D_h$, and (c) $x = 18D_h$.}\label{fig:12}
\end{figure}

The time variations of the wall temperatures of the microchannels with different heat fluxes are presented in Figure \ref{fig:12}. As the bubbles flow through the cross-section of interests periodically, the wall temperature fluctuates strongly. With the increase of $q_w$, the magnitude of temperature fluctuation increases. This is because with increasing $q_w$, the bubbles absorb more heat by vaporization and become larger (also see Figure \ref{fig:10}), then the larger bubble sizes further affect the heat transfer with the wall and also affect the local wall temperature. It should be noted that the frequency of the temperature fluctuation is not affected by the heat fluxes even though the bubbles are larger. This is because temperature fluctuation frequency is determined by the bubble formation frequency, which is fixed in the upstream. 

\subsection{Effect of latent heat of the working fluid}\label{sec:3.3}
\begin{figure}
  \centering
  \includegraphics[width=\columnwidth]{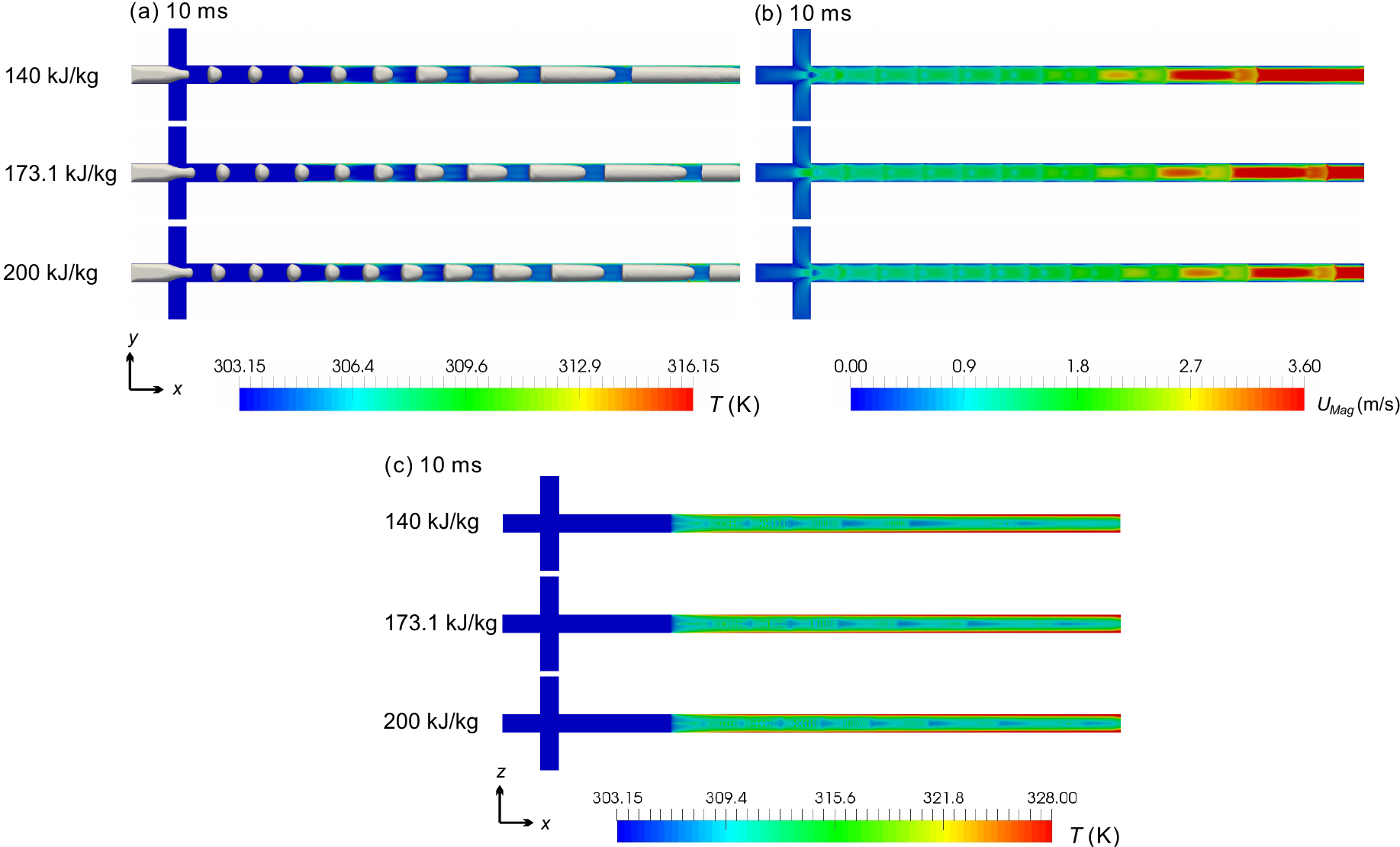}
  \caption{Bubble train in microchannel flow boiling for different latent heat at a typical instant of the stabilized state: (a) bubble shapes and temperature fields in the middle cross-section in the $z$-direction; (b) velocity field at middle cross-section in the $z$-direction; (c) bottom wall temperature illustrating the temperature distribution of the heated wall for different latent heat in the stabilized state.}\label{fig:13}
\end{figure}

The latent heat of the working fluid has a significant impact on the efficiency of phase change application systems. The understanding of the effect of latent heat in microchannels can lead to improved designs of micro-device cooling systems. The CFD approach allows us to isolate the influence of latent heat on the flow and heat transfer characteristics without the practical complexities of physically modifying the fluid. In this study, the effect of different latent heat of vaporization of the working fluid on the flow dynamics of the bubble train is studied during microchannel flow boiling. The bubble train temperature field and flow field are presented in Figure \ref{fig:13} for different latent heat at a typical instant of the stabilized state. The latent heat has a significant effect on the flow dynamics and the heat transfer of the bubble train as shown in Figure \ref{fig:13}(a). With increasing the latent heat, the bubbles' growth rate decreases. This is because with the same amount of wall heat absorption by the fluid, the rate of vaporization reduces with increasing the latent heat, hence inhibiting the expansion of the vapor bubbles. As a consequence, with increasing the latent heat, the bubbles in the microchannels become small, and allow more bubbles to exist at the same time in the microchannel, as shown in Figure \ref{fig:13}(a). In addition, because of the slower growth rate of the bubbles at higher latent heat, the acceleration effect of the bubble is inhibited and the flow speed in the channel becomes smaller, as shown in Figure \ref{fig:13}(b). The wall temperature distribution at different latent heat is shown in Figure \ref{fig:13}(c). With increasing the latent heat, the temperature variation along the channel has a higher frequency. This is because of the smaller bubble size. The latent heat does not have a significant impact on the overall temperature on the bottom wall. Although the fluid with higher latent heat has the capacity to absorb more heat, the vaporization rate also becomes smaller. Therefore, the overall temperature magnitude is insensitive to the latent heat.

\begin{figure}
  \centering
  \includegraphics[scale=0.27]{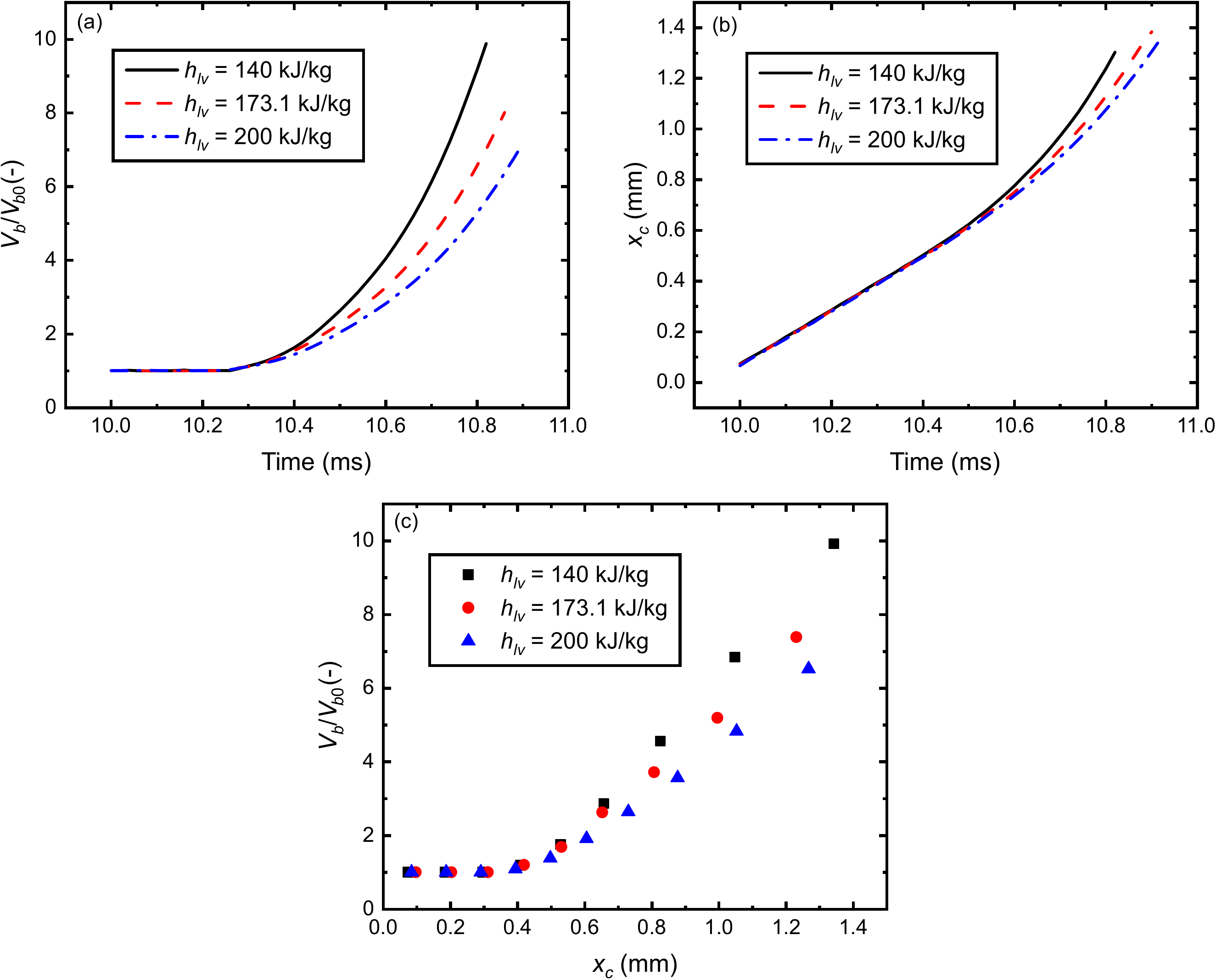}
  \caption{Effects of latent heat of the working fluid on the volume and position of the bubbles in the stabilized state: (a) Time variation of the dimensionless bubble volume $V_b/V_{b0}$ of a single bubble; (b) Time variation of the dimensionless bubble position $x_c/D_h$ of a single bubble; (c) Dimensionless bubble volume $V_b/V_{b0}$ versus bubble center position at a typical instant.}\label{fig:14}
\end{figure}

The effect of the latent heat on the bubble expansion is quantified by the volume and the position of the bubbles, as shown in Figure \ref{fig:14}. In the adiabatic region, the bubble size is unchanged and the bubbles moves at a constant speed, as shown in Figures \ref{fig:14}(a) and \ref{fig:14}(b). As the bubbles enters the heated region, the bubble volume quickly increases because of the heat absorption from the wall and the vaporization of the fluid. Meanwhile, the bubble accelerates because of the bubble expansion. With increasing the latent heat, the vaporization rate becomes slower. Hence, the bubble growth rate decreases and the bubble acceleration becomes smaller, as shown in Figure \ref{fig:14}(a) and \ref{fig:14}(b) respectively. At a typical instant, the volumes of the bubbles in the microchannels are plotted against the bubble position in Figure \ref{fig:14}(c). We can see that the bubbles in the downstream is much larger than that in the upstream, and the bubbles of fluids with higher latent heat are much smaller than that of lower latent heat, which is consistent with the bubble volume in Figure \ref{fig:14}(a). 

\subsection{Capillary effect on the bubble train ($Ca$)}\label{sec:3.4}
\begin{figure}
  \centering
  \includegraphics[width=\columnwidth]{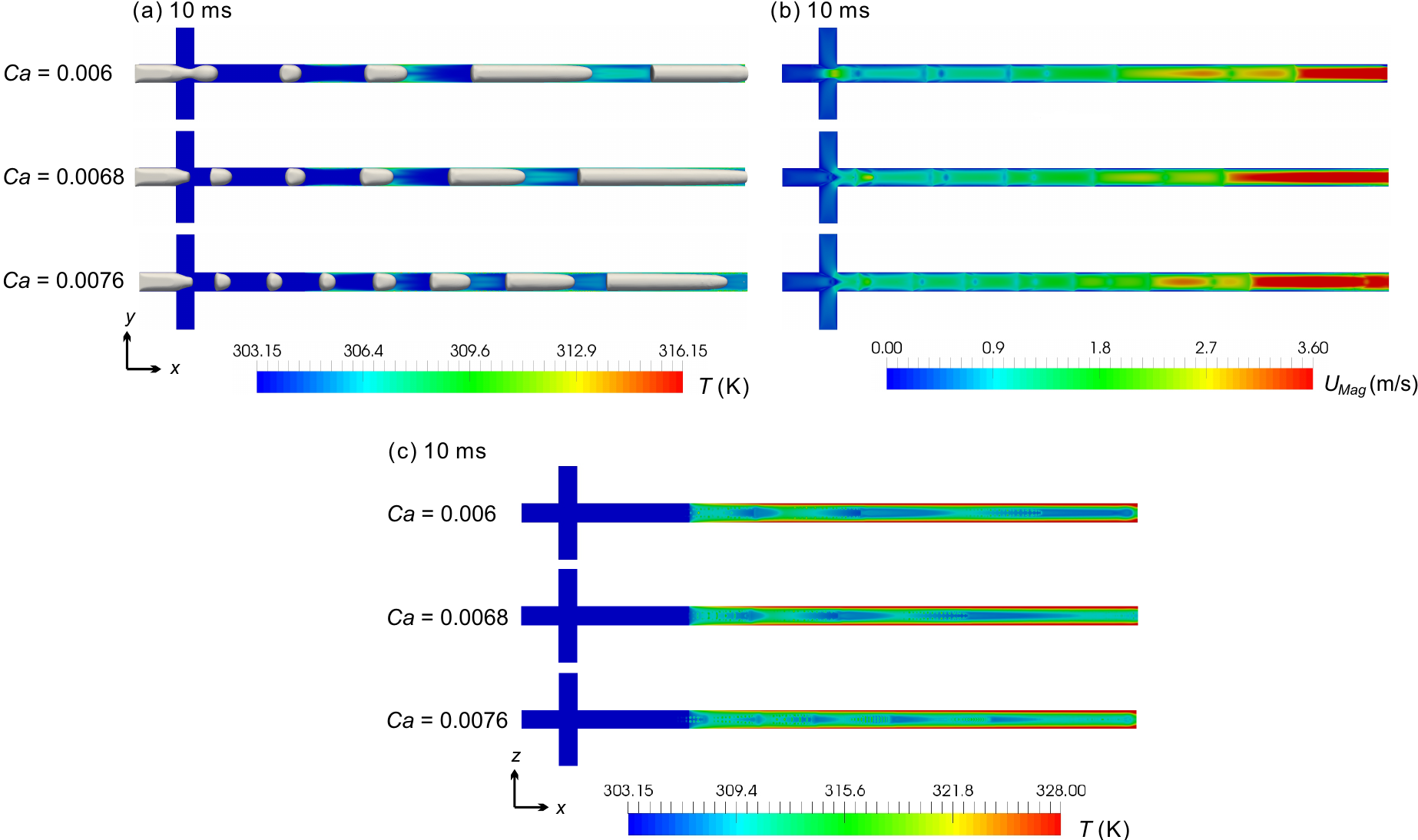}
  \caption{Bubble train in microchannel flow boiling for different $Ca$ at a typical instant of the stabilized state: (a) bubble shapes and temperature fields in the middle cross-section in the $z$-direction; (b) velocity field at middle cross-section in the $z$-direction; (c) bottom wall temperature illustrating the temperature distribution of the heated wall for different $Ca$ in the stabilized state.}\label{fig:15}
\end{figure}

The capillary number, which quantifies the ratio between the viscous force and surface tension force, is an important parameter for the flow boiling process. To understand the capillary effects on the flow boiling process, the capillary number $Ca = \mu_l U_l /\sigma$ is varied by changing the liquid inlet velocity, while fixing the ratio of the vapor/liquid inlet velocities to $U_v/U_l = 0.5$ (corresponding to the initial vapor-liquid volume ratio of $VR = 0.25$). As shown in Figure \ref{fig:15}, the capillary number not only affects the bubble formation at the flow focusing junction, but also affects the bubble growth and heat transfer in the downstream.

As the $Ca$ number increases, the bubble frequency increases and bubble size decreases, as shown in Figure \ref{fig:15}(a). This is because a higher $Ca$ number indicates a higher effect of viscous force and a weaker effect of the surface tension force. The viscous force acts as the driving force for the bubble breakup, and the surface tension force acts as the inhibitive force. Therefore, the bubble becomes easier to break up as the $Ca$ number increases, producing more bubbles of smaller sizes in the channel. In the downstream of the channel, as the $Ca$ number increases, the bubbles tend to be more significantly deformed by the flow. As bubbles move in microchannels, they generally adopt bullet shapes because of the competition between the surface tension force and the viscous force. The viscous force tends to deform the bubbles because of the large velocity gradient in the microchannels, while the surface tension force tends to restore the bubbles to spherical by minimizing the interfacial energy. As the $Ca$ number increases, the viscous force increases, which elongates the bubbles. The elongation of the bubbles increases the liquid film beneath the bubbles, hence affecting the heat transfer process. The case with a higher $Ca$ number has a thicker liquid film, which leads to slower bubble growth than that of a lower $Ca$ number with faster and larger bubble growth due to its closer contact with the superheated thermal boundary layer.

The flow field in the microchannel is shown in Figure \ref{fig:15}(b). As the $Ca$ number increases, although we have higher inlet velocity at higher $Ca$ numbers, the velocity difference between different $Ca$ numbers is not significant. This is because the velocity in the downstream is significantly affected by the growth of the vapor bubbles in the upstream, which accelerate the liquid in the downstream to a higher velocity. At a higher $Ca$ number (corresponding to higher inlet velocity), because of the deformation of the bubble shape and slower bubble growth, the acceleration to the downstream fluid by the bubble growth in the upstream is less. The effects of the inlet velocity and the effect of the acceleration by the upstream bubble growth on the downstream velocity can cancel each other to some extent. Hence, the velocity in the downstream between different $Ca$ numbers becomes insignificant. As a consequence of the velocity field, the effect of the $Ca$ number on the wall temperature is also insignificant, as shown in Figure \ref{fig:15}(c).

\begin{figure}
  \centering
  \includegraphics[scale=0.51]{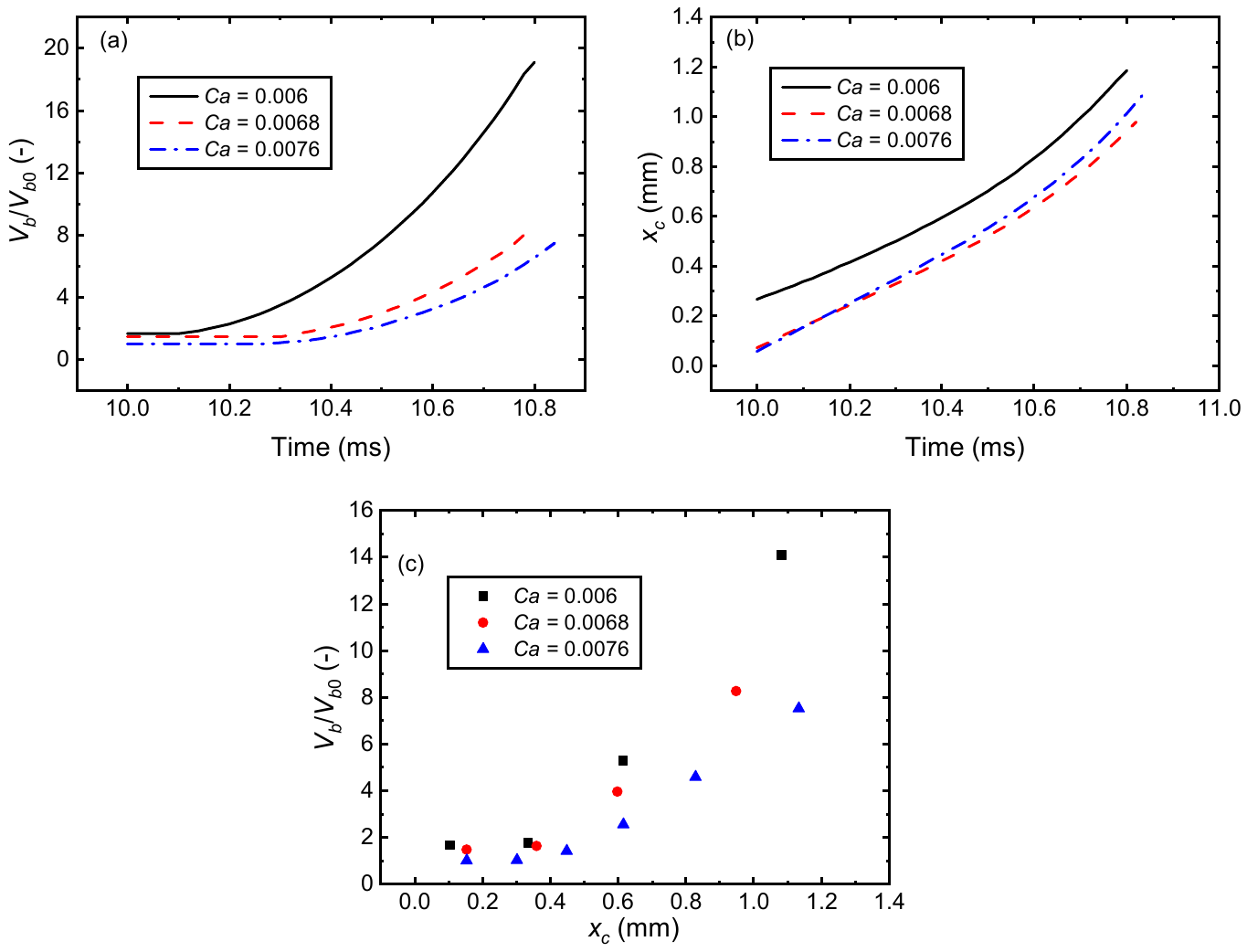}
  \caption{Capillary effects on the volume and position of the bubbles in the stabilized state: (a) Time variation of the dimensionless bubble volume $V_b/V_{b0}$ of a single bubble; (b) Time variation of the dimensionless bubble position $x_c/D_h$ of a single bubble; (c) Dimensionless bubble volume $V_b/V_{b0}$ versus bubble center position at a typical instant.}\label{fig:16}
\end{figure}

To analyze the bubbles' growth in the microchannel at different $Ca$ numbers, Figure \ref{fig:16}(a) shows the time variation of the dimensionless bubble volume $V_b/V_{b0}$ for a single bubble in the bubble train in the stabilized state. In the adiabatic region, the bubble size generated at lower $Ca$ numbers is larger and grows faster than those at higher $Ca$ numbers due to its closer contact with the superheated thermal boundary layer, as discussed in Figure \ref{fig:15}. The dimensionless bubble volumes $V_b/V_{b0}$ at a typical instant are shown in Figure \ref{fig:16}(c), with each symbol representing a bubble in the bubble train and consistent with the bubble size variation over time. The bubble position versus time at different $Ca$ numbers is shown in Figure \ref{fig:16}(b). In the upstream, the bubble at the higher $Ca$ number has a higher speed. However, in the downstream, the bubble speed is significantly affected by the bubble growth in the upstream, which accelerates the bubble movements in the downstream. Therefore, at higher $Ca$ numbers, the effect of higher inlet velocity and the effect of slower bubble growth can cancel each other to some extent. Therefore, the influence of the capillary number on the bubble movement is insignificant, which is consistent with the results presented in Figure \ref{fig:16}(b).

\section{Conclusions}\label{sec:4}
In this study, the dynamics of bubble train in microchannel flow boiling heat transfer is explored. We adopt the flow-focusing technique to achieve continuous vapor bubble trains in microchannels with fine-tuned bubble size and frequency, and focus on the bubble dynamics and heat transfer in the stabilized state. The effects of different vapor-liquid volume ratios, different heat fluxes, and different latent heat on the flow dynamics of the bubble train and the heat transfer are investigated. Such stabilized state is important for microchannel flow boiling heat transfer, but was not captured in previous studies, which only simulated single isolated bubbles or a few bubbles. The effects of different vapor-liquid volume ratios, different heat fluxes, different latent heat, and different Capillary number on the flow dynamics of the bubble train and the heat transfer are investigated. The main conclusions of this study are summarized as follows:
	
\begin{enumerate}
  \item With increasing the vapor-liquid volume ratio, the bubble frequency increases while the growth rate of the bubble train decreases. However, by increasing the bubble frequency, the heat transfer is improved.
  \item As the heat flux increases, the growth rate of the bubble train increases because of the increased vaporization rate.
  \item With increasing the latent heat, the growth rate of the bubble train decreases because of the slower vaporization rate.
  \item As the Capillary number increases, the bubble frequency increases while the growth rate of the bubble train decreases because the increased viscous force promotes the breakup of bubbles. The increased Capillary number also increases the thickness of the liquid film beneath the bubble, which inhibits the bubble growth.
  \item The speeds of the bubbles are not only affected by the bubble size, but also affected by the vaporization of the bubbles in the upstream, which pushes the bubbles in the downstream forwards.
  \item The wall temperature fluctuates as the bubbles pass through periodically in the microchannels. The temperature fluctuation is influenced by the initial vapor-liquid volume ratio, the wall heat flux, the fluid latent heat, and the Capillary number. The fluctuation of the wall temperature also leads to the fluctuation of the Nusselt number.
\end{enumerate}

This study not only provide physical insight into microchannel flow boiling processes, but are also helpful for the applications of microchannel heat sinks. 

\section*{Declaration of Competing Interest}
None.

\section*{Acknowledgements}
This work was supported by the National Natural Science Foundation of China (Grant Nos.\ 51920105010 and 51921004).

\bibliographystyle{elsarticle-num}
\bibliography{BubbleTrain}

\begin{thebibliography}{10}
\expandafter\ifx\csname url\endcsname\relax
  \def\url#1{\texttt{#1}}\fi
\expandafter\ifx\csname urlprefix\endcsname\relax\def\urlprefix{URL }\fi
\expandafter\ifx\csname href\endcsname\relax
  \def\href#1#2{#2} \def\path#1{#1}\fi

\bibitem{karayiannis17}
T.~G. Karayiannis, M.~M. Mahmoud, Flow boiling in microchannels: Fundamentals
  and applications, Applied Thermal Engineering 115 (2017) 1372--1397.
\newblock \href {https://doi.org/10.1016/j.applthermaleng.2016.08.063}
  {\path{doi:10.1016/j.applthermaleng.2016.08.063}}.

\bibitem{mudawar13}
I.~Mudawar, Recent advances in high-flux, two-phase thermal management, Journal
  of Thermal Science and Engineering Applications 5~(2) (2013) 021012.
\newblock \href {https://doi.org/10.1115/1.4023599}
  {\path{doi:10.1115/1.4023599}}.

\bibitem{valehesheyda13}
P.~Valeh-e Sheyda, M.~Rahimi, E.~Karimi, M.~Asadi, Application of two-phase
  flow for cooling of hybrid microchannel {PV} cells: A comparative study,
  Energy Conversion and Management 69 (2013) 122--130.
\newblock \href {https://doi.org/10.1016/j.enconman.2013.01.029}
  {\path{doi:10.1016/j.enconman.2013.01.029}}.

\bibitem{wu24}
H.~Wu, S.~Zhou, D.~Wang, Y.~Yang, L.~Liu, H.~Mao, B.~Shu, Predictive modeling
  for microchannel flow boiling heat transfer under the dual effect of gravity
  and surface modification, Processes 12~(5) (2024) 1028.
\newblock \href {https://doi.org/10.3390/pr12051028}
  {\path{doi:10.3390/pr12051028}}.

\bibitem{mudawar11}
I.~Mudawar, Two-phase microchannel heat sinks: theory, applications, and
  limitations, Journal of Electronic Packaging 133~(4) (2011) 041002.
\newblock \href {https://doi.org/10.1115/1.4005300}
  {\path{doi:10.1115/1.4005300}}.

\bibitem{thome04}
J.~R. Thome, Boiling in microchannels: a review of experiment and theory,
  International Journal of Heat and Fluid Flow 25~(2) (2004) 128--139.
\newblock \href {https://doi.org/10.1016/j.ijheatfluidflow.2003.11.005}
  {\path{doi:10.1016/j.ijheatfluidflow.2003.11.005}}.

\bibitem{Huh2007FlowBoiling}
C.~Huh, C.~W. Choi, M.~H. Kim, Elongated bubble behavior during flow boiling in
  a microchannel, Journal of Mechanical Science and Technology 21~(11) (2007)
  1819.
\newblock \href {https://doi.org/10.1007/BF03177438}
  {\path{doi:10.1007/BF03177438}}.

\bibitem{yan23}
S.-l. Yan, X.-q. Wang, L.-t. Zhu, X.-b. Zhang, Z.-h. Luo, Mechanisms and
  modeling of bubble dynamic behaviors and mass transfer under gravity: A
  review, Chemical Engineering Science 277 (2023) 118854.
\newblock \href {https://doi.org/10.1016/j.ces.2023.118854}
  {\path{doi:10.1016/j.ces.2023.118854}}.

\bibitem{harirchian11}
T.~Harirchian, S.~V. Garimella, Boiling heat transfer and flow regimes in
  microchannels—a comprehensive understanding, Journal of Electronic
  Packaging 133~(1) (2011) 011001.
\newblock \href {https://doi.org/10.1115/1.4002721}
  {\path{doi:10.1115/1.4002721}}.

\bibitem{thome15}
J.~R. Thome, A.~Cioncolini, Two-phase flow pattern maps for microchannels,
  World Scientific, 2015, pp. 47--84.
\newblock \href {https://doi.org/10.1142/9789814623216_0020}
  {\path{doi:10.1142/9789814623216_0020}}.

\bibitem{cheng17}
L.~Cheng, G.~Xia, Fundamental issues, mechanisms and models of flow boiling
  heat transfer in microscale channels, International Journal of Heat and Mass
  Transfer 108 (2017) 97--127.
\newblock \href {https://doi.org/10.1016/j.ijheatmasstransfer.2016.12.003}
  {\path{doi:10.1016/j.ijheatmasstransfer.2016.12.003}}.

\bibitem{kadam21}
S.~T. Kadam, I.~Hassan, R.~Kumar, M.~A. Rahman, Bubble dynamics in
  microchannel: An overview of the state-of-the-art, Meccanica 56~(3) (2021)
  481--513.
\newblock \href {https://doi.org/10.1007/s11012-020-01300-4}
  {\path{doi:10.1007/s11012-020-01300-4}}.

\bibitem{Revellin2006DiabaticTwoPhaseFlow}
R.~Revellin, V.~Dupont, T.~Ursenbacher, J.~R. Thome, I.~Zun, Characterization
  of diabatic two-phase flows in microchannels: Flow parameter results for
  {R-134a} in a 0.5mm channel, International Journal of Multiphase Flow 32~(7)
  (2006) 755--774.
\newblock \href {https://doi.org/10.1016/j.ijmultiphaseflow.2006.02.016}
  {\path{doi:10.1016/j.ijmultiphaseflow.2006.02.016}}.

\bibitem{Mukherjee2005BubbleGrowth}
A.~Mukherjee, S.~G. Kandlikar, Numerical simulation of growth of a vapor bubble
  during flow boiling of water in a microchannel, Microfluidics and
  Nanofluidics 1~(2) (2005) 137--145.
\newblock \href {https://doi.org/10.1007/s10404-004-0021-8}
  {\path{doi:10.1007/s10404-004-0021-8}}.

\bibitem{mukherjee11}
A.~Mukherjee, S.~G. Kandlikar, Z.~J. Edel, Numerical study of bubble growth and
  wall heat transfer during flow boiling in a microchannel, International
  Journal of Heat and Mass Transfer 54~(15-16) (2011) 3702--3718.
\newblock \href {https://doi.org/10.1016/j.ijheatmasstransfer.2011.01.030}
  {\path{doi:10.1016/j.ijheatmasstransfer.2011.01.030}}.

\bibitem{Zhuan2012FlowBoilingMicrochannel}
R.~Zhuan, W.~Wang, Flow pattern of boiling in micro-channel by numerical
  simulation, International Journal of Heat and Mass Transfer 55~(5) (2012)
  1741--1753.
\newblock \href {https://doi.org/10.1016/j.ijheatmasstransfer.2011.11.029}
  {\path{doi:10.1016/j.ijheatmasstransfer.2011.11.029}}.

\bibitem{Ferrari2018SlugFlowBoiling}
A.~Ferrari, M.~Magnini, J.~R. Thome, Numerical analysis of slug flow boiling in
  square microchannels, International Journal of Heat and Mass Transfer 123
  (2018) 928--944.
\newblock \href {https://doi.org/10.1016/j.ijheatmasstransfer.2018.03.012}
  {\path{doi:10.1016/j.ijheatmasstransfer.2018.03.012}}.

\bibitem{luo20}
Y.~Luo, W.~Li, K.~Zhou, K.~Sheng, S.~Shao, Z.~Zhang, J.~Du, W.~J. Minkowycz,
  Three-dimensional numerical simulation of saturated annular flow boiling in a
  narrow rectangular microchannel, International Journal of Heat and Mass
  Transfer 149 (2020) 119246.
\newblock \href {https://doi.org/10.1016/j.ijheatmasstransfer.2019.119246}
  {\path{doi:10.1016/j.ijheatmasstransfer.2019.119246}}.

\bibitem{guo16}
Z.~Guo, B.~S. Haynes, D.~F. Fletcher, Numerical simulation of annular flow
  boiling in microchannels, The Journal of Computational Multiphase Flows 8~(1)
  (2016) 61--82.
\newblock \href {https://doi.org/10.1177/1757482X16634205}
  {\path{doi:10.1177/1757482X16634205}}.

\bibitem{priy24}
A.~Priy, I.~Ahmad, M.~K. Khan, M.~Pathak, Bubble interaction and heat transfer
  characteristics of microchannel flow boiling with single and multiple
  cavities, Journal of Thermal Science and Engineering Applications 16~(6)
  (2024) 061010.
\newblock \href {https://doi.org/10.1115/1.4065187}
  {\path{doi:10.1115/1.4065187}}.

\bibitem{zhang24}
X.~Zhang, Y.~Zhang, C.~Yang, J.~Liu, B.~Liu, Z.~Zhou, Study of flow and heat
  transfer performance of nanofluids in curved microchannels with different
  curvatures, Numerical Heat Transfer, Part A: Applications (2024) 1--22\href
  {https://doi.org/10.1080/10407782.2024.2321520}
  {\path{doi:10.1080/10407782.2024.2321520}}.

\bibitem{odumosu23}
O.~A. Odumosu, H.~Xu, T.~Wang, Z.~Che, Growth of elongated vapor bubbles during
  flow boiling heat transfer in wavy microchannels, Applied Thermal Engineering
  223 (2023) 119987.
\newblock \href {https://doi.org/10.1016/j.applthermaleng.2023.119987}
  {\path{doi:10.1016/j.applthermaleng.2023.119987}}.

\bibitem{zhang23}
Z.~Zhang, G.~Zhang, M.~Wei, Y.~Zhang, M.~Tian, Numerical simulation of boiling
  behavior in vertical microchannels, Physics of Fluids 35~(9) (2023) 092011.
\newblock \href {https://doi.org/10.1063/5.0167304}
  {\path{doi:10.1063/5.0167304}}.

\bibitem{rajkotwala22}
A.~H. Rajkotwala, L.~L. Boer, E.~A. J.~F. Peters, C.~W.~M. van~der Geld,
  J.~G.~M. Kuerten, J.~A.~M. Kuipers, M.~W. Baltussen, A numerical study of
  flow boiling in a microchannel using the local front reconstruction method,
  AIChE Journal 68~(4) (2022) e17598.
\newblock \href {https://doi.org/10.1002/aic.17598}
  {\path{doi:10.1002/aic.17598}}.

\bibitem{Liu2017BubbleTrainMicrochannel}
Q.~Liu, W.~Wang, B.~Palm, C.~Wang, X.~Jiang, On the dynamics and heat transfer
  of bubble train in micro-channel flow boiling, International Communications
  in Heat and Mass Transfer 87 (2017) 198--203.
\newblock \href {https://doi.org/10.1016/j.icheatmasstransfer.2017.07.002}
  {\path{doi:10.1016/j.icheatmasstransfer.2017.07.002}}.

\bibitem{Magnini2016FlowBoiling}
M.~Magnini, J.~R. Thome, Computational study of saturated flow boiling within a
  microchannel in the slug flow regime, Journal of Heat Transfer 138~(2) (2016)
  021502.
\newblock \href {https://doi.org/10.1115/1.4031234}
  {\path{doi:10.1115/1.4031234}}.

\bibitem{bertsch09}
S.~S. Bertsch, E.~A. Groll, S.~V. Garimella, Effects of heat flux, mass flux,
  vapor quality, and saturation temperature on flow boiling heat transfer in
  microchannels, International Journal of Multiphase Flow 35~(2) (2009)
  142--154.
\newblock \href {https://doi.org/10.1016/j.ijmultiphaseflow.2008.10.004}
  {\path{doi:10.1016/j.ijmultiphaseflow.2008.10.004}}.

\bibitem{zhu17}
P.~Zhu, L.~Wang, Passive and active droplet generation with microfluidics: a
  review, Lab on a Chip 17~(1) (2017) 34--75.
\newblock \href {https://doi.org/10.1039/C6LC01018K}
  {\path{doi:10.1039/C6LC01018K}}.

\bibitem{baroud10}
C.~N. Baroud, F.~Gallaire, R.~Dangla, Dynamics of microfluidic droplets, Lab on
  a Chip 10~(16) (2010) 2032--2045.
\newblock \href {https://doi.org/10.1039/C001191F}
  {\path{doi:10.1039/C001191F}}.

\bibitem{scheufler23}
H.~Scheufler, J.~Roenby, {TwoPhaseFlow}: A framework for developing two phase
  flow solvers in {OpenFOAM}, OpenFOAM® Journal 3 (2023) 200--224.
\newblock \href {https://doi.org/10.51560/ofj.v3.80}
  {\path{doi:10.51560/ofj.v3.80}}.

\bibitem{Brackbill1992ModelingSurfaceTension}
J.~U. Brackbill, D.~B. Kothe, C.~Zemach, A continuum method for modeling
  surface tension, Journal of Computational Physics 100~(2) (1992) 335--354.
\newblock \href {https://doi.org/10.1016/0021-9991(92)90240-y}
  {\path{doi:10.1016/0021-9991(92)90240-y}}.

\bibitem{hardt08}
S.~Hardt, F.~Wondra, Evaporation model for interfacial flows based on a
  continuum-field representation of the source terms, Journal of Computational
  Physics 227~(11) (2008) 5871--5895.
\newblock \href {https://doi.org/10.1016/j.jcp.2008.02.020}
  {\path{doi:10.1016/j.jcp.2008.02.020}}.

\bibitem{Luo2017NumericalBubbleGrowth}
Y.~Luo, J.~Zhang, W.~Li, E.~Sokolova, Y.~Li, W.~Minkowycz, Numerical
  investigation of the bubble growth in horizontal rectangular microchannels,
  Numerical Heat Transfer, Part A: Applications 71~(12) (2017) 1175--1188.
\newblock \href {https://doi.org/10.1080/10407782.2017.1350538}
  {\path{doi:10.1080/10407782.2017.1350538}}.

\bibitem{Anna2016DropletsBubbles}
S.~L. Anna, Droplets and bubbles in microfluidic devices, Annual Review of
  Fluid Mechanics 48 (2016) 285--309.
\newblock \href {https://doi.org/10.1146/annurev-fluid-122414-034425}
  {\path{doi:10.1146/annurev-fluid-122414-034425}}.

\bibitem{Anna2003FormationDispersions}
S.~L. Anna, N.~Bontoux, H.~A. Stone, Formation of dispersions using ``flow
  focusing'' in microchannels, Applied Physics Letters 82~(3) (2003) 364--366.
\newblock \href {https://doi.org/10.1063/1.1537519}
  {\path{doi:10.1063/1.1537519}}.

\bibitem{Moragues2023DropletMicrofluidics}
T.~Moragues, D.~Arguijo, T.~Beneyton, C.~Modavi, K.~Simutis, A.~R. Abate, J.-C.
  Baret, A.~J. deMello, D.~Densmore, A.~D. Griffiths, Droplet-based
  microfluidics, Nature Reviews Methods Primers 3~(1) (2023) 32.
\newblock \href {https://doi.org/10.1038/s43586-023-00212-3}
  {\path{doi:10.1038/s43586-023-00212-3}}.

\bibitem{che20153d}
Z.~Che, T.~N. Wong, N.-T. Nguyen, C.~Yang, Three dimensional features of
  convective heat transfer in droplet-based microchannel heat sinks,
  International Journal of Heat and Mass Transfer 86 (2015) 455--464.
\newblock \href {https://doi.org/10.1016/j.ijheatmasstransfer.2015.03.030}
  {\path{doi:10.1016/j.ijheatmasstransfer.2015.03.030}}.

\end{thebibliography}

\end{document}